\documentclass[pra,aps,twocolumn,showpacs]{revtex4-2}
\usepackage{amsmath,amssymb,graphicx}
\usepackage{float}
\usepackage{comment}
\usepackage{bm}
\usepackage{bbold}

\usepackage{graphicx}
\usepackage{dcolumn}
\usepackage{bm}
\usepackage{hyperref}
\usepackage{xcolor}
\usepackage{braket}
\usepackage{textcomp}
\usepackage{comment}

\newcommand{\om}[1]{\Omega_{#1\tilde{#1}}}
\newcommand{\oms}[2]{\Omega_{#2\tilde{#1}}}

\newcommand{\g}[1]{g_{#1\tilde{#1}}}
\newcommand{\gc}[1]{\bar{g}_{#1\tilde{#1}}}
\newcommand{\gs}[2]{g_{#2\tilde{#1}}}

 \newcommand{\s}[3]{\sigma_{#1 #2}^{#3}}
 
\newcommand{\kk}{\mathbf{k}}

\newcommand{\rr}{\mathbf{r}}

\begin{document}

\preprint{APS/123-QED}

\title{Tunable photon-mediated interactions between spin-1 systems}

\author{Cristian Tabares}
\affiliation{%
Institute of Fundamental Physics IFF-CSIC, Calle Serrano 113b, 28006 Madrid, Spain}%
\author{Erez Zohar}
\address{Racah Institute of Physics, Hebrew University
of Jerusalem, Jerusalem 91904 , Israel}
\author{Alejandro González-Tudela}%
 \email{a.gonzalez.tudela@csic.es}
\affiliation{%
Institute of Fundamental Physics IFF-CSIC, Calle Serrano 113b, 28006 Madrid, Spain.}%

\date{\today}

\begin{abstract}
The exchange of virtual photons between quantum optical emitters in cavity QED or quantum nanophotonic setups induces interactions between them which can be harnessed for quantum information and simulation purposes. So far, these interactions have been mostly characterized for two-level emitters, which restrict their application to engineering quantum gates among qubits or simulating spin-1/2 quantum many-body models. Here, we show how to harness multi-level emitters with several optical transitions to engineer a wide class of photon-mediated interactions between effective spin-1 systems. We characterize their performance through analytical and numerical techniques, and provide specific implementations based on the atomic level structure of Alkali atoms. Our results expand the quantum simulation toolbox available in such cavity QED and quantum nanophotonic setups, and open up new ways of engineering entangling gates among qutrits.
\end{abstract}

\maketitle


\section{Introduction} \label{sec:introduction}

Non-local quantum correlations are the key resource of most quantum information and simulation technologies~\cite{Altman2021,Awschalom2021,Alexeev2021}. One way of obtaining them between quantum emitters is through the exchange of photons via their optical transitions~\cite{lehmberg70a,lehmberg70b}. When such exchange involves mostly off-resonant (virtual) photons, like it occurs in the dispersive regime of cavity QED~\cite{ritsch13a,Aron2016,vaidya18a,welte18a,Bentsen2019,Periwal2021} or in the ``band-gap regime" of quantum nanophotonic platforms~\cite{douglas15a,Gonzalez-Tudela2015b,Hung2016,hood16a,Chang2018,evans18a,Samutpraphoot2020}, it induces coherent photon-mediated interactions between the emitters which can be harnessed for engineering entangling gates~\cite{Duan2005,Lin2006,welte18a,Samutpraphoot2020} or simulating exotic many-body Hamiltonians~\cite{douglas15a,Gonzalez-Tudela2015b,Hung2016,Bentsen2019}, among other applications. Since these interactions can be longer-ranged than in other platforms, they provide a way of observing novel many-body phases, such as supersolid~\cite{Leonard2017,Bottcher2019,Chomaz2019}, magnetic~\cite{Barnett2006,Micheli2006,maik12a,Glaetzle2015}, or topological~\cite{Manmana2013,Yao2013,Bello2019a,Bello2022} ones, difficult to obtain otherwise. 

Remarkably, except for a few works~\cite{Birnbaum2006,Norris2012,Kurucz2010,Terraciano2009,Hamley2012,Zhiqiang2017,landini18a,Morales2019,Davis2019,Davis2020,Kohler2018,Norris2010,Arnold2011,Hemmer2021,norcia18a,Masson2017}, most of the studies have focused so far on characterizing these interactions and their consequences between (effective) two-level systems. This sets limitations, for example, on the types of gates that can be engineered, i.e., only qubit ones, and the many-body models that can be simulated, i.e., spin 1/2 systems. Since quantum emitters, like atomic systems, can have a much richer level structure, there is an increasing interest in the last few years~\cite{Xu2021,Campos-Gonzalez-Angulo2021,Groiseau2021,Orioli2021} on harnessing it to find exotic phenomena, such as multi-critical behaviour in Dicke phase transitions~\cite{Xu2021} or emergent dark entangled states~\cite{Orioli2021}, as well as to develop new applications, such as new multi-photon sources~\cite{Orioli2021}. One very attractive reason for considering multi-level emitters is the possibility to engineer photon-mediated interactions between higher-dimensional spins, which can find applications in the quantum simulation of non-trivial high-energy physics problems~\cite{zohar2015,Dalmonte2016,Banuls2020,Zohar2022,Aidelsburger2022,Klco2021}, to prepare symmetry-protected topological states in spin-1 chains~\cite{Affleck1987,Kairys2022}, to solve complex optimization problems~\cite{Deller2022}, and, more generally, to engineer universal quantum gates between spin-1 systems~\cite{Wang2020a}.

Here, we show how to harness multi-level emitters to engineer different types of spin-1 photon-mediated interactions (ZZ and XX) in cavity QED and quantum nanophotonic setups. For that, we use a combination of judiciously chosen Raman-assisted transitions which connect the ground and excited state levels that, after tracing out the photonic and excited-state degrees of freedom, result in different photon-mediated interactions between the effective spin-1 system appearing in their ground state manifold. To characterize them, we use projection operator techniques for open quantum systems~\cite{Reiter2012} to find the effective dynamics and characterize the performance of the interactions as entangling gates. We do our analysis in two steps: first, in a platform-agnostic way, so that our results can be of interest to different type of multi-level emitters (such as quantum dots, vacancy centers or atoms), and then, particularizing for the multi-level structure of a particular atom, i.e., Rubidium. In the latter case, we will fully take into account the different Clebsch-Gordan coefficients of the transitions, and explain how to compensate the corrections introduced by them.  The text is structured as follows: in Section~\ref{sec:system}, we explain the general setup and theoretical framework that we consider along the manuscript; in Section~\ref{sec:general} we analyze in a platform-agnostic way the different type of photon-interactions that can be obtained in these setups; Then, in Section~\ref{sec:atoms} we particularize for an atomic system, taking into account the complexity introduced by the Clebsch-Gordan coefficients; in Section~\ref{sec:applications}, we enumerate a few examples where such spin-1 photon mediated interactions can be exploited, and finally in Section~\ref{sec:conclu} we summarize our main findings and conclude.
\begin{figure*}[t]
    \centering
    \includegraphics[width=0.7\linewidth]{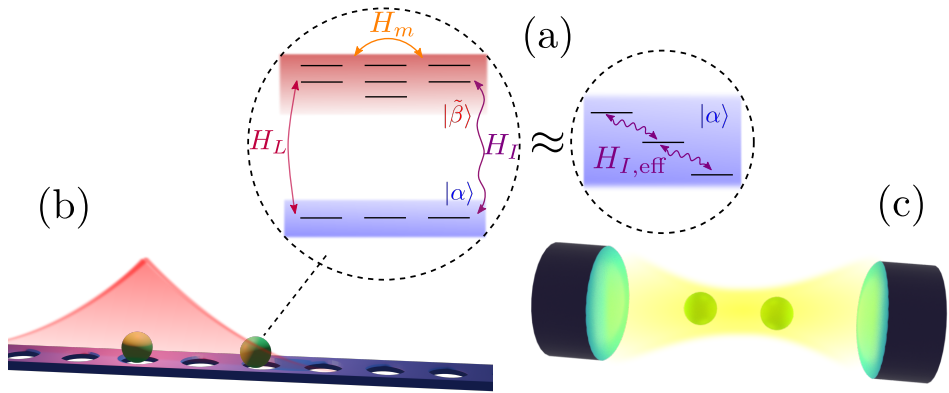}
    \caption{(a) Left: full multilevel structure of the quantum emitters considered, differentiating between ground/slow (blue) and excited/fast (red) state subspaces. We also depict the Raman lasers connecting them (represented by a Hamiltonian $H_L$ in the main text), the coupling between the emitter and the quantized cavity/waveguide field (denoted by $H_I$), and also the additional microwave driving generating transitions between the ground/excited subspaces ($H_m$). Right: in the conditions where the excited states can be adiabatically eliminated, see the main text for details, the evolution of the system can be captured by an effective light-matter Hamiltonian $H_{I,\text{eff}}$ with only transitions between the ground state levels. If the photon field can further be eliminated, one then obtains effective photon-mediated interactions between the ground state levels of the different emitters. (b-c) Schemes of the two photonic setups of interest: (b) Two emitters coupled to a nanophotonic waveguide. When the emitter's frequency lie within a band-gap, their emission becomes localized forming what has been called as atom-photon bound-state~\cite{john90a,kurizki90a} (in red), which can mediate purely coherent interactions~\cite{douglas15a,Gonzalez-Tudela2015b,Hung2016}  between the emitters. (c) Two emitters trapped inside an optical cavity. When the emitter's optical transition are far off-resonance from the cavity one, purely coherent interactions between emitters are also generated.}
    \label{fig:1}
\end{figure*}
\section{System and theoretical framework~\label{sec:system}}

The general setup that we consider along this manuscript is sketched in Fig.~\ref{fig:1}: several multi-level quantum optical emitters (Fig~\ref{fig:1}a) are interfaced with the photonic modes confined in a photonic waveguide (Fig~\ref{fig:1}b) or cavity (Fig~\ref{fig:1}c). The multi-level structure of the emitters is comprised by a ground (with slow dynamics) and an excited (with faster ones) state manifold denoted by: $\left\{\ket{\alpha}\right\}=\left\{\ket{1},\ket{2},\dots\right\}$ and $\left\{\ket{\tilde{\beta}}\right\}=\left\{\ket{\tilde{1}},\ket{\tilde{2}},\dots\right\}$, respectively, and whose intrinsic Hamiltonians are $H_{\text{s/f}}$, with $H_{\text{s(f)}}=\sum_i\sum_{\alpha(\tilde{\beta})}\omega_{\alpha(\tilde{\beta})}\sigma^i_{\alpha\alpha(\tilde{\beta}\tilde{\beta})}$, where we use the notation $\sigma^i_{\alpha\tilde{\beta}}=\ket{\alpha}_i\bra{\tilde{\beta}}$ for the atomic operators of the $i$-th atom (we take $\hbar=1$ throughout this manuscript).

The ground and excited state manifolds are connected by two different mechanisms:
\begin{itemize}

\item Either through classical laser fields described by Hamiltonians:
\begin{equation}
        H_L(t) =\sum_i \sum_{\alpha,\tilde{\beta}}\frac{\Omega_{\alpha\tilde{\beta}}^{i}}{2}\sigma_{\alpha\tilde{\beta}}^{i}e^{i\omega_{\alpha\tilde{\beta}}t}+\text{H.c.},\label{eq:HL}
\end{equation}
with $\Omega_{\alpha\tilde{\beta}}^{i}$ and $\omega_{\alpha\tilde{\beta}}$ being the amplitude and frequency of the laser driving the $\alpha\leftrightarrow\tilde{\beta}$ transition of the $i$-th emitter (in general we will take the same amplitude for each emitter, $\Omega_{\alpha\tilde{\beta}}^{i}\equiv \Omega_{\alpha\tilde{\beta}}$).

\item Or via photon exchange with the cavity/nanophotonic confined modes. We can describe both situations through the following light-matter interaction Hamiltonian:
\begin{equation}
        H_I = \sum_i\sum_{\alpha,\tilde{\beta}}g_{\alpha\tilde{\beta}}^{i}\sigma_{\alpha\tilde{\beta}}^{i} A_i^\dagger +\text{H.c.},\label{eq:LM}
    \end{equation}
where $g_{\alpha\tilde{\beta}}^{i}$ is the coupling strength of the $\alpha\leftrightarrow\tilde{\beta}$-optical transition of the $i$-th atom to the photonic mode $A_i^\dagger$, which would be $A_i^\dagger=a^\dagger$ for the single-mode cavity QED setups if all emitters couple to equivalent positions of the cavity, or $A_i^\dagger=\sum_\kk e^{-i\kk\cdot\rr_i}a_\kk^\dagger$ for the case where the emitters interact with the continuum of photon modes confined in a nanophotonic structure. This light-matter Hamiltonian has to be complemented by the one associated with the energy of the photonic modes, which reads $H_a=\omega_a a^\dagger a$ or $H_a=\sum_{\kk}\omega(\kk)a_\kk^\dagger a_\kk$, for the cavity and nanophotonic situation, respectively.
    
\end{itemize}

On top of that, we assume that there can be additional couplings within the excited or ground state manifolds, e.g., using microwave drivings, described by:
 \begin{equation}
        H_m (t)= \sum_{i}\sum_{\tilde{\beta},\tilde{\beta}'} \frac{\Omega^{i}_{\tilde{\beta}\tilde{\beta}'}}{2}\sigma_{\tilde{\beta}\tilde{\beta}'}^i e^{i\omega_{\tilde{\beta}\tilde{\beta}'}t}+\mathrm{H.c.}, 
    \end{equation}

Thus, the complete dynamics of the emitters and bath system is given by a Hamiltonian with all the previously described contributions, i.e., $H(t)=H_\text{s}+H_\text{f}+H_a+H_L(t)+H_I+H_m(t)$. However, both the cavity and excited state levels are generally subject to losses. This means that the system must be described by a density matrix, $\rho(t)$, whose dynamics is governed by the following Born-Markov master equation:
\begin{align}
\label{eq:theory_master}
    \frac{\mathrm{d}}{\mathrm{d}t}\rho(t) = i\left[\rho,H(t)\right]+\sum_j\left( L_j \rho L_j^\dagger-\frac{1}{2}\left(L_j^\dagger L_j \rho + \rho L_j^\dagger L_j\right)\right),
\end{align}
where $L_j$ are the different jump operators describing the noise processes. For example, $L_{\kappa}=\sqrt{\kappa}a$ accounts for the cavity losses at rate $\kappa$, whereas $L_{\gamma_{\alpha\tilde{\beta}}}=\sqrt{\gamma_{\alpha\tilde{\beta}}}\sigma_{\alpha\tilde{\beta}}$ accounts for the incoherent decay between the excited state $\tilde{\beta}$ and the ground state $\alpha$ at rate $\gamma_{\alpha\tilde{\beta}}$.

In this work, we are interested in the photon-mediated interactions appearing in effective spin-1 systems arising in the ground-state manifold. To obtain their shape, we will adiabatically eliminate the other degrees of freedom, that are, excited-state and photonic ones, using the projection operator techniques for open quantum systems developed in Ref.~\cite{Reiter2012}. These adiabatic elimination techniques can be applied when there is a separation of timescales within the Hilbert space in which some of states evolve much slower than the rest. Then, one can define projection operators over the fast/slow subspaces, denoted by $\mathbb{P}_\mathrm{f}$ and $\mathbb{P}_\mathrm{s}$, respectively, satisfying $\mathbb{P}_\mathrm{s}+\mathbb{P}_\mathrm{f}=\mathbb{1}$ and $\mathbb{P}_\mathrm{f}\mathbb{P}_\mathrm{s}=\mathbb{P}_\mathrm{s}\mathbb{P}_\mathrm{f}=0$. Using these operators, we can describe the interactions inside the ground (excited) subspace with $H_\mathrm{s}\equiv \mathbb{P}_\mathrm{s}H\mathbb{P}_\mathrm{s}$ ($H_\mathrm{f}\equiv \mathbb{P}_\mathrm{f}H\mathbb{P}_\mathrm{f}$), and the connections between them with $V_+ \equiv \mathbb{P}_\mathrm{f}H\mathbb{P}_\mathrm{s}$ ($V_-\equiv \mathbb{P}_\mathrm{s}H\mathbb{P}_\mathrm{f}=V_+^\dagger$), with which one can obtain an effective master equation for the slow subspace~\cite{Reiter2012}:
\begin{eqnarray}\label{eq:theory_master_eff}
    \frac{\mathrm{d}}{\mathrm{d}t}&&\rho(t) = i\left[\rho,H_\text{eff}\right]\\
    &&+\sum_j L_\text{eff}^j \rho (L_\text{eff}^j)^\dagger-\frac{1}{2}\left[(L_\text{eff}^j)^\dagger L_\text{eff}^j \rho + \rho (L_\text{eff}^j)^\dagger L_\text{eff}^j\right]\,,\nonumber
\end{eqnarray}
where we have introduced the effective Hamiltonian,
\begin{equation}\label{eq:theory_effH}
    H_\text{eff} = H_\mathrm{s}-\frac{1}{2}V_- \left[H_\mathrm{NH}^{-1}+(H_\mathrm{NH}^{-1})^\dagger\right]V_+\,,
\end{equation}
and the effective Linblad operators $L_\text{eff}^j = L_j H_\mathrm{NH}^{-1}V_+$, defined in terms of the non-Hermitian Hamiltonian used in the quantum jump formalism,
\begin{equation}\label{eq:HNH}
    H_\mathrm{NH} = H_\mathrm{f}-\frac{i}{2}\sum_j L_j^\dagger L_j\,.
\end{equation}

Looking at Eq.~\eqref{eq:theory_effH}, it  can be seen that the accuracy of this effective evolution will increase with the energy gap between the ground and excited subspaces but will decrease as the amplitude of the perturbative (de-) excitations $V_\pm$ grows. Finally, let us emphasize again that these expressions have been obtained in the conditions that $||H_s||\ll ||H_\mathrm{NH}||$, such that the evolution within the slow subspace is neglected when compared with the fast one.
\section{General analysis of emergent spin-1 photon-mediated interactions~\label{sec:general}}

 In this Section we will apply the previously described formalism in two steps: first, we will eliminate the emitter excited-state manifold to obtain an effective light-matter Hamiltonian as in Eq.~\eqref{eq:LM}, but with renormalized parameters depending on the laser configuration. Then, we will eliminate the photonic field under the Born-Markov assumptions~\cite{breuer-petruccione} to obtain the effective photon-mediated interactions between spin-1 systems. We illustrate it with two minimal examples of increasing complexity and show how one can obtain effective ZZ and XX interactions in these systems. For concreteness, we will only do the derivations for the cavity QED setup, i.e., $A^\dagger_i=a^\dagger$ in Eq.~\eqref{eq:LM}, although the expressions can be readily generalized to the nanophotonic setups in the band-gap regime, as explained in subsection~\ref{subsec:nano} and shown with more detail in Refs.~\cite{douglas15a,Gonzalez-Tudela2015b,Hung2016}. Besides, let us note that in this section we will just explain the minimal level structure required to obtain the different type of interactions, so that our findings can be of interest for different types of emitters (quantum dots, vacancy centers, atoms,\dots). Then, in the next Section~\ref{sec:atoms}, we will explain how to obtain these configurations with Alkali atoms.
 
\subsection{ZZ interactions~\label{subsec:ZZabstract}}

\subsubsection{Hamiltonian dynamics~\label{subsubsec:ZZcoherent}}
\begin{figure}[tb]
    \centering
    \includegraphics[width=\linewidth]{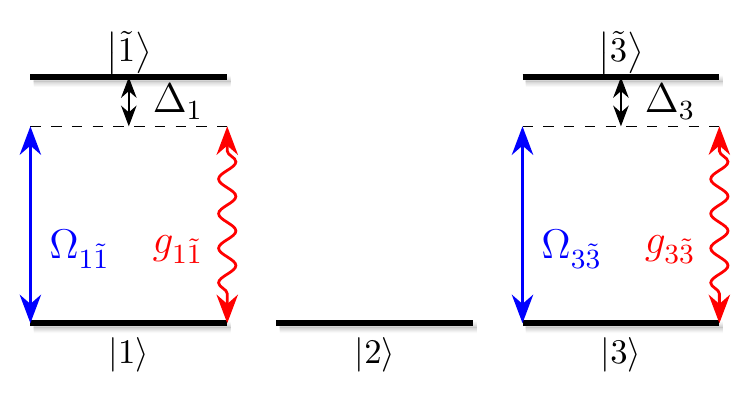}
\caption{Minimal multi-level configuration to obtain an spin-1 Ising ZZ interaction ground state manifold: one requires three ground-state levels, $1,2,3$, to codify a spin-1 operator. Then, two of them must be connected through a laser (blue)/cavity (red) field to an optically excited state as depicted in the picture.}
    \label{fig:int_zz_scheme}
\end{figure}

Let us first illustrate the method explaining how to obtain the simplest instance of spin-1 interaction, that is, the Ising ZZ interaction, $S^i_z S_z^j$, being $S^i_z$ the spin-1 Z operator of the $i$-th atom. Here, we will discuss first the effective Hamiltonian evolution obtained, and then consider the non-Hermitian terms in subsection~\ref{subsubsec:ZZlosses}). The minimal level structure required is depicted in Fig.~\ref{fig:int_zz_scheme}: one needs at least three ground-state levels ($1,2,3$) to codify the spin-1 operators, that with our notation can be written as: $S_z^i = \s{1}{1}{i}-\s{3}{3}{i}$. To engineer the non-local gates between emitters, one can connect the states $1$ and $3$ to two optically excited states, $\tilde{1}$ and $\tilde{3}$, as depicted in Fig.~\ref{fig:int_zz_scheme}, with laser fields, and design the cavity structure so that it can also connect back to $1$ and $3$. 

Going to a rotating frame oscillating at the laser frequency, $\omega_L$, of the lasers $\Omega_{1\tilde{1}}$ and $\Omega_{3\tilde{3}}$, the Hamiltonian part of the fast subspace (excited state levels plus cavity) reads:
\begin{equation}\label{eq:int_zz_ex}
    H_\text{f}=\sum_{i}\left(\Delta_1\s{\tilde{1}}{\tilde{1}}{i}+\Delta_2 \s{\tilde{2}}{\tilde{2}}{i} + \Delta_{3}\s{\tilde{3}}{\tilde{3}}{i}\right)+\Delta_a a^\dagger a\,,
\end{equation}
where $\Delta_i=\omega_i-\omega_L$, with $i=1,2,3,a$. The Hamiltonians connecting the slow (ground-state manifold described by a Hamiltonian $H_\text{s}$) and fast subspaces in this case are given by both the laser fields:
\begin{equation}
    H_L = \sum_{i}\left(\frac{\om{1}}{2}\s{\tilde{1}}{1}{i}+\frac{\om{3}}{2}\s{\tilde{3}}{3}{i}+\mathrm{H.c.}\right)\,,
\end{equation}
and the light-matter Hamiltonian which in this case reads:
\begin{equation}\label{eq:int_zz_I}
    H_I = \sum_{i}\left(\g{1}^i \s{1}{\tilde{1}}{i}a^\dagger+\g{3}^i\s{3}{\tilde{3}}{i}a^\dagger+\mathrm{H.c.}\right)\,.
\end{equation}

Note, we intentionally assume that neither the cavity mode nor the laser field couple the $2$ state with any other excited state, which is critical to obtain the right effective Hamiltonian. In Section~\ref{sec:atoms}, we provide a way on how this can be achieved using optical selection rules, e.g., in Rubidium.

With that separation between the different Hamiltonian terms, and assuming that the laser amplitudes and cavity couplings  are far smaller than their detunings, $\mathrm{max}|\Omega_{\alpha\tilde{\beta}}|\ll \mathrm{max}\left|\Delta_i\right|$ ($i=1,2,3$) and $\mathrm{max}|g_{\alpha\tilde{\beta}}|\ll \mathrm{max}\left|\Delta_a\right|$ so that there is a timescale separation between the two subspaces, we can apply the projection operator technique explained in the previous section, see Eq.~\ref{eq:theory_master_eff}, to adiabatically eliminate the excited-state and photonic degrees of freedom to obtain (the details of the derivation can be found in the Appendix~\ref{app:derivations}):
\begin{eqnarray}\label{eq:zz_eff}
    H_\mathrm{eff} &&= H_\text{s}-\sum_{i}\Bigg[\left[\frac{\Delta_1|\om{1}|^2}{4\Delta_1^2 + \gamma_1^2}\s{1}{1}{i}+\frac{\Delta_3|\om{3}|^2}{4\Delta_3^2 + \gamma_3^2}\s{3}{3}{i}\right]\nonumber\\
    &&+\frac{4\Delta_a}{4\Delta_a^2+\kappa^2}\left[|\mu_i|^2\s{1}{1}{i}+|\nu_i|^2\s{3}{3}{i}\right]\Bigg]+H_\text{int},
\end{eqnarray}
with $H_\text{int}$ being:
\begin{eqnarray}
    H_\text{int}=-\frac{4\Delta_a}{4\Delta_a^2+\kappa^2}\sum_{i\neq j}&&\mu_i\bar{\mu}_j\s{1}{1}{i}\s{1}{1}{j}+\nu_i\bar{\nu}_j\s{3}{3}{i}\s{3}{3}{j}\\&&+\left(\mu_i\bar{\nu_j}\s{1}{1}{i}\s{3}{3}{j}+\bar{\mu_i}\nu_j\s{1}{1}{j}\s{3}{3}{i}\right)\,,\nonumber\label{eq:zz_eff2}
\end{eqnarray}
and where we have introduced the parameters:
\begin{equation}\label{eq:munu_def}
        \mu_i = \frac{2\om{1}\gc{1}^i\Delta_1}{4\Delta_1^2+\gamma_1^2},\,\,\text{and}\,\,\nu_i = \frac{2\om{3}\gc{3}^i\Delta_3}{4\Delta_3^2+\gamma_3^2},
\end{equation}
to simplify the notation (the overbar means complex conjugate). Note that, although only Hamiltonian dynamics have been considered up until now, the non-Hermitian contribution coming from the photon-loss processes for both the excited-state levels, $L_{\gamma_\alpha} = \sum_i \sqrt{\gamma_{\alpha}}\s{\alpha}{\tilde{\alpha}}{i}$, and cavity modes, $L_\kappa = \sqrt{\kappa}a$ already enter these formulas, renormalizing the detunings. Furthermore, as expected, we obtain two different terms: on the one hand, the energies of the levels $1,3$ become shifted by the interactions with the off-resonant laser/cavity fields, as captured by Eq.~\ref{eq:zz_eff}. On the other hand, the exchange of photons between the atoms leads to a non-local exchange between different atoms, see Eq.~\ref{eq:zz_eff2}, whose particular shape depends on the atomic configuration chosen ($\Omega$'s,$g$'s,$\Delta$'s). For example, by choosing the parameters such that $\mu_i=-\nu_i$, the state-dependent shifts of Eq.~\ref{eq:zz_eff} become equal for both levels. Thus, they can be removed connecting both of them to another far detuned excited state to generate an additional AC Stark shift of the same absolute value and opposite sign, so that they oscillate with the same phase than level $2$. Another option is to induce a similar shift on the $2$-level by connecting it with another independent excited state. Irrespective of the method chosen, the effective dynamics of the ground-state manifold can be mapped to a pure Ising spin-1 ZZ Hamiltonian:
\begin{equation}\label{eq:eff_ising}
   H_\text{eff} = H_\text{s} + \sum_{i<j} J_\mathrm{z}^{ij} S^{i}_z S^{j}_z\,,
\end{equation}
with
\begin{equation}\label{eq:eff_ising_Jz}
  J_{\mathrm{z}}^{ij} = \Re\left[\frac{-8\Delta_a\mu_i \bar{\mu}_j }{4\Delta_a^2+\kappa^2}\right]\,.
\end{equation}

\subsubsection{Non-unitary contributions~\label{subsubsec:ZZlosses}}

\begin{figure*}[t]
    \centering
    \includegraphics[width=\textwidth]{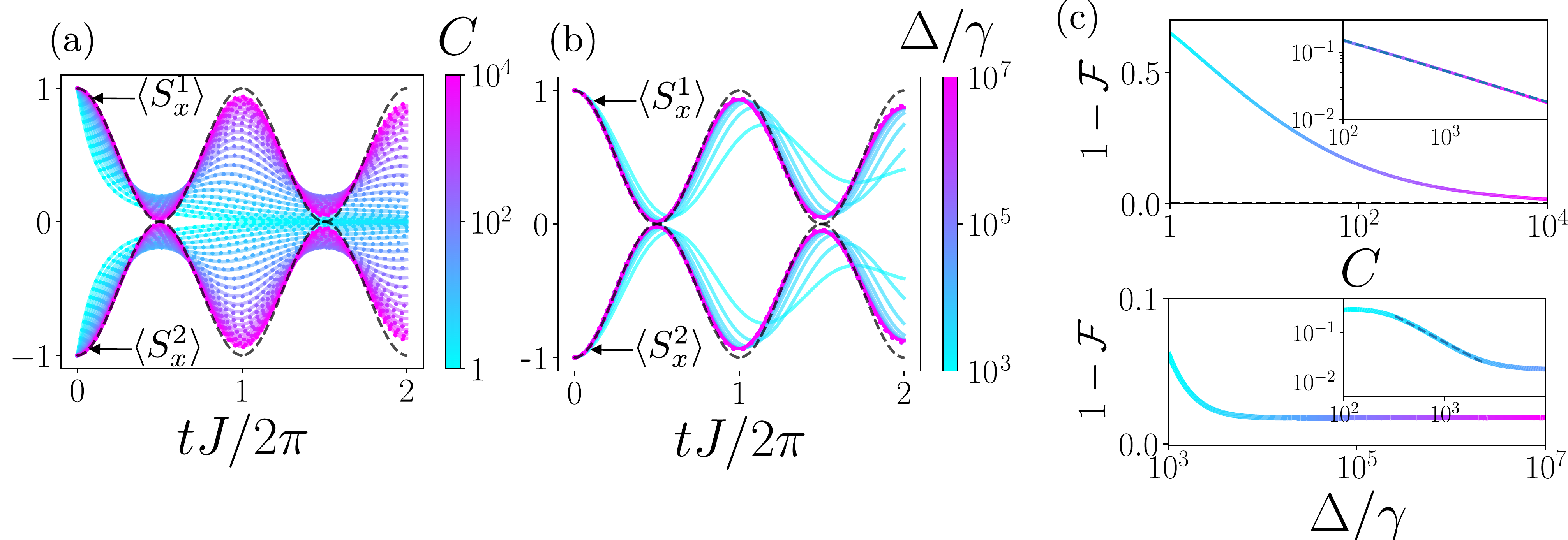}
    \caption{(a) Mean values of the $\langle S_{x}^{i}\rangle$ operators as functions of time for two multilevel quantum emitters interacting with a quantized mode and driven with an external field, according to the scheme in Fig.~\ref{fig:int_zz_scheme} and including losses, for different values of the cooperativity $C=g^2/(\kappa\gamma)$. The solid lines show the numerical calculation obtained with a full evolution, considering the emitters' slow and fast subspaces and also the field mode, meanwhile the dots correspond to the effective operators obtained in Eqs.~\ref{eq:eff_ising} and~\ref{eq:losses_ising} (for the coherent part and the losses, respectively) already projected onto the slow subspace. We note that we take $H_\text{s}=0$, assuming we already remove the AC Stark shifts from the levels. The initial state was $\ket{\Psi(0)}=\ket{m=1}_{x}\otimes \ket{m=-1}_x$, and the parameters were $\Delta_1 = \Delta_3 := \Delta$, $\Omega_1 = \Omega_3 = \Delta/20$, $\Delta_a = -100\Delta$, $g_3=-g_1=\sqrt{\gamma/\kappa}\Delta_a$ (so condition~\ref{eq:optimal_det} is satisfied and the detuning $\Delta_a$ minimizes the losses), $\gamma$ ranges from $\Delta/10^{5}$ to $\Delta/10^{3}$ and $\gamma/\kappa = 10^{-5}$ to keep $\epsilon_\text{approx}$ constant and low (the dashed lines show the ideal evolution, with $\gamma,\kappa\rightarrow0$ and therefore $C\rightarrow\infty$). (b) Same as in (a), but keeping the cooperativity constant at $C=10^{4}$ and instead modifying the value of the rate of spontaneous emission against the detuning, $\Delta/\gamma$, meanwhile $\kappa=|\Delta^{\text{opt}}_{a}|/\sqrt{C}$ so condition~\ref{eq:optimal_det} is again satisfied. Note that all the effective evolutions collapse onto the same dots. (c) Top panel: Infidelity (defined as $\mathcal{I}=1-\mathcal{F}$, Eq.~\ref{eq:infidelity_def}) where the fidelity $\mathcal{F}$ (Eq.~\ref{eq:fidelity_def}) quantifies the overlap between the state $\rho_\mathrm{target}$ at $t=\pi/J$ ($J:=J^{ii}_{\text{z}}$, Eq.~\ref{eq:eff_ising_Jz}) obtained with a fully coherent evolution according to the effective Hamiltonian in Eq.~\ref{eq:eff_ising} and the actual state obtained after a real evolution of the system including the full Hamiltonian with the excited levels and also the non-Hermitian contributions, taking the same parameters as in (a). The inset shows the same plot in a log-log scale, with a numerical line fit of slope $\sim 0.46$ superimposed, in agreement with the expected scaling of $\epsilon_{\text{losses}}$. Bottom panel: Same as in the top panel, but now using the parameters of (b) and plotting the infidelity as a function of $\Delta/\gamma$. The inset shows a numerical line fit of slope $\sim -1.09$ superimposed, also in agreement with the expected scaling of $\epsilon_{\text{approx}}$.}
    \label{fig:zz_agnostic}
\end{figure*}

In the previous subsection we considered only the effective Hamiltonian dynamics induced. However, let us note that together with the effective Hermitian Hamiltonian in Eq.~\eqref{eq:eff_ising}, both the cavity and atomic decay rates, $\kappa$ and $\gamma_{i}$'s respectively, induce non-unitary dynamics that will compete with the effective unitary dynamics induced by the photon-mediated interactions. In particular, the terms corresponding to spontaneous decay would lead to the operators
\begin{equation}
    L^\text{eff}_{\gamma_j} =\sum_{i} \frac{\om{j}\sqrt{\gamma_j}}{2\Delta_j -i \gamma_j}\s{j}{j}{i},
\end{equation}
with $\gamma_j$ the corresponding decay rate and $j=1,3$ (for simplicity we will take $\gamma_j :=\gamma$ for all $j$). Meanwhile the loss of photons will induce terms of the form:
\begin{eqnarray}
    L^\text{eff}_\kappa = \frac{2\sqrt{\kappa}}{2\Delta_a-i\kappa}\sum_i \left(\mu_i \s{1}{1}{i}+\nu_i \s{3}{3}{i}\right).
\end{eqnarray}

To compare the effects of these terms with the effective coherent interaction given by Eq.~\ref{eq:eff_ising}, we can use the non-Hermitian Hamiltonian of the quantum jump formalism with the effective interaction and jump operators found, respectively. We focus on the case where $\mu_i = -\nu_i$, which is the situation of interest for this manuscript, assuming that all the parameters ($\Omega$'s, $g$'s, $\Delta$'s and $\gamma$'s) are equal in absolute value for all the levels. In that scenario, the non-Hermitian terms entering the effective Hamiltonian to take into account the losses would be
\begin{eqnarray}\label{eq:losses_ising}
    \sum_j L^{\text{eff }\dagger}_{\gamma_j} L^\text{eff}_{\gamma_j} = \frac{|\Omega|^2 \gamma}{4\Delta^2+\gamma^2}\sum_i S_{z}^{i\,2}\quad\text{and}\nonumber\\
    L^{\text{eff }\dagger}_{\kappa} L^\text{eff}_{\kappa} = \sum_i \frac{4\kappa|\mu_i|^2}{4\Delta_a^2+\kappa^2}S_{z}^{i\,2}=\sum_i \frac{J_{\mathrm{z}}^{ii}\kappa}{2\Delta_a}S_{z}^{i\,2}.
\end{eqnarray}

Therefore, in the complete evolution of the system, there are two competing processes: a coherent evolution according to the effective Hamiltonian in Eq.~\eqref{eq:eff_ising}, and the non-unitary dynamics induced by the incoherent terms given by Eq.~\eqref{eq:losses_ising}. From the first one we find that the time needed to appreciate the effects of the effective Hamiltonian is of order $\tau\sim 1/J_{\text{z}}^{ij}$. Thus, the probability of inducing an error in the coherent evolution in that period of time due to the losses would be of order:
\begin{equation}
    \epsilon_{\text{losses}} \sim \tau \left[\frac{|\Omega|^2 \gamma}{4\Delta^2+\gamma^2}+\frac{4\kappa|\mu_i|^2}{4\Delta_a^2+\kappa^2}\right],
\end{equation}
that becomes:
\begin{equation}\label{eq:epsilon_losses}
    \epsilon_{\text{losses}} \sim \frac{4\Delta_a}{g^2}\left[\gamma + \kappa\left(\frac{g}{\Delta_a}\right)^2\right]
\end{equation}
to first order in $\gamma/\Delta$ and $\kappa/\Delta_a$. This expression shows the expected trade-off between cancelling spontaneous emission or cavity decay errors that occur in other cavity QED situations with simple two-level emitters~\cite{douglas15a}. To reduce this value, we may optimize $\epsilon_\text{losses}$ as a function of $\Delta_a$ to find an optimal detuning: 
\begin{equation}\label{eq:optimal_det}
    \Delta^{\text{opt}}_a =g\sqrt{\frac{\kappa}{\gamma}}\,,
\end{equation}
that yields an error scaling as $\epsilon_\text{losses}\sim 1/\sqrt{C}$, where $C=g^2/(\kappa\gamma)$ is the single-atom cooperativity \cite{douglas15a}. 

However, apart from the errors coming from spontaneous emission and photon decays, there might appear additional ones due to deviations from the adiabatic elimination conditions. In particular, it is necessary that the effective slow subspace dynamics (with energy scale $\sim J_{\text{z}}^{ij}$) are much slower than the fast subspace ones (with corresponding energy scale $\Delta\sim\Delta_{\text{a}}$). We can then parametrize the errors introduced by the projection onto the slow subspace as  $\epsilon_\text{approx} \sim J^{ij}_{\text{z}}/\Delta$, so the condition of different timescales means that $\epsilon_{\text{approx}}\ll 1$.  Imposing the optimal detuning in Eq.~\eqref{eq:optimal_det} and expanding the result up to first order in $\gamma/\Delta$ and $\kappa/\Delta_a$, we find the scaling:
\begin{equation}\label{eq:epsilon_approx}
    \epsilon_\text{approx} \sim \frac{\left|\Omega\right|^2}{\Delta^2} \frac{|g|^2}{\Delta^{\text{opt}}_{a}\Delta}=\frac{\left|\Omega\right|^2}{\Delta^2}\frac{\gamma}{\kappa}\frac{\Delta_{a}^{\text{opt}}}{\Delta}=\frac{\left|\Omega\right|^2}{\Delta^2}\frac{\gamma}{\Delta}\sqrt{C}\,.
\end{equation}

Thus, when comparing the expected quantum evolution that could be found in a real setup (hence including the excited levels and also the non-Hermitian jump operators) with the ideal one, given only by the effective coherent part of the photon-mediated interactions, the total error (that is, the infidelity $\mathcal{I}$ between the states) will be such as:
\begin{equation}\label{eq:infidelity_def}
    \mathcal{I} = 1-\mathcal{F} \approx \epsilon_{\text{losses}}+\epsilon_{\text{approx}}+\dots\,,
\end{equation}
where the dots include other possible terms of higher order or origin that we do not consider in this discussion, and where the fidelity $\mathcal{F}$ is defined as
\begin{equation}\label{eq:fidelity_def}
    \mathcal{F}=\mathrm{Tr}\left(\sqrt{\sqrt{\rho}\rho_\mathrm{target}\sqrt{\rho}}\right)^2\,.
\end{equation}
This quantity quantifies the overlap between the state $\rho_\text{target}$ obtained only with the effective Hamiltonian (that is, a fully coherent evolution) and the actual $\rho$ including the non-Hermitian contributions and also the excited levels. 

Therefore, according to Eq.~\eqref{eq:infidelity_def}, the cooperativity cannot be increased indefinitely in order to completely reduce the errors, since this will decrease $\epsilon_{\text{losses}}$ but  will increase $\epsilon_{\text{approx}}$. If we fix the ratio $\Omega/\Delta$ (which sets a general timescale in the dynamics and should not be excessively small, or the dynamical evolutions would be quite slow), and aim to reduce $\epsilon_{\text{losses}}$ increasing the cooperativity, one needs also to keep the ratio $\gamma/\Delta$ low (or conversely, $\Delta/\gamma$ high) in order to keep $\epsilon_{\text{approx}}$ small. 

Let us now numerically benchmark the accuracy of all these expressions and arguments by comparing the evolution obtained using the effective operators against the exact, full quantum evolution. We acknowledge that all the numerical calculations shown in this manuscript have been performed with the aid of the QuTip package~\cite{Johansson2013}. This comparison is what we show in Fig.~\ref{fig:zz_agnostic}. There, we initialize two emitters in the state $\ket{\Psi(0)}=\ket{m=1}_x \otimes \ket{m=-1}_x$ and let them evolve with the full Hamiltonian (solid lines) $H= H_\mathrm{f}+H_L+H_I$ (Eqs.~\eqref{eq:int_zz_ex}-\eqref{eq:int_zz_I}) and the effective Hamiltonian shown in Eq.~\eqref{eq:eff_ising} (dotted lines). In both cases we include the possible losses of atoms and photons, properly projected onto the slow subspace for the effective situation, and take $H_\text{s}=0$, i.e., assuming that we compensate the different start shifts between the ground state levels. First, in Fig.~\ref{fig:zz_agnostic}(a), we check how the dynamics governed by the effective Hamiltonian (dotted) deviates from the exact one (solid) for system with different cooperativities $C$, and a small ratio $\gamma/\Delta$ so that the error introduced by the projection operator technique, $\epsilon_\text{approx}$, is low. As expected, high values of the cooperativity show a behaviour much similar to the ideal case (perfect oscillations with period $T=2\pi/J$, $J:= J_{\text{z}}^{ii}$, shown with black dashed lines), meanwhile the simulations with $C\sim 1$ show almost no coherent dynamics because everything is governed by the non-unitary processes. In Fig.~\ref{fig:zz_agnostic}(b) we fix the cooperativity to a large value, $C=10^4$, and move the ratio $\Delta/\gamma$, so that it explores situations where $\epsilon_\text{approx}$ becomes larger. There, we see how for the smaller values of $\Delta/\gamma\sim 10^3$, the dynamics of the effective Hamiltonian (dotted) deviate significantly from the exact ones (solid). Note that in this case the errors do not appear because of a predominance of the non-unitary processes, but rather deviations from the purely coherent dynamics. Thus, they induce systematic errors that, in certain cases, could be corrected by choosing appropriately gate times.

To give a quantitative estimation of the accuracy in both cases, we also show in Fig.~\ref{fig:zz_agnostic}(c) the infidelity, Eq.~\eqref{eq:infidelity_def}, to quantify the total error between the Hermitian effective evolution and the actual one including the losses. In both cases we fix $t_\text{target}=\pi/J$ and use the parameters in Figs.~\ref{fig:zz_agnostic}(a) and \ref{fig:zz_agnostic}(b) for the top and bottom panels, respectively. The top panel shows indeed that the expected error decrease as the cooperativity is increased as long as $\gamma/\Delta\ll 1$, so the contribution from $\epsilon_{\text{approx}}$ can be neglected. Under these conditions, $\mathcal{I}\approx \epsilon_{\text{losses}}$, and we find a numerical scaling of $\mathcal{I}\sim 1/C^{0.46}$ (Fig.~\ref{fig:zz_agnostic}c, inset on the top panel) in approximate agreement with the $\epsilon_\text{losses}\sim 1/\sqrt{C}$ discussed above (the slope tends to $0.5$ as bigger values $\Delta/\gamma$ are considered, since this reduces the effect of the other contributions to the error). On the other hand, the bottom panel of Fig.~\ref{fig:zz_agnostic}(c) confirms that keeping a high value of the cooperativity ($C=10^4$) is not enough to obtain an accurate effective evolution, since the actual dynamics including the whole Hilbert space can be different if $\Delta/\gamma$ is not big enough. In particular, we see that the total error decreases as $\Delta/\gamma$ is increased up to a constant value (which is given by the error due to the losses, $\epsilon_{\text{losses}}$, that is fixed by the finite value of the cooperativity), and the linear fit to the loglog plot around this region shown in the inset of the bottom panel of Fig.~\ref{fig:zz_agnostic}(c) yields a scaling of $\mathcal{I}\approx \epsilon_{\text{approx}}\sim \left(\Delta/\gamma\right)^{-1.09}$, in agreement with Eq.~\ref{eq:epsilon_approx} (a more detailed study of these arguments can be found in Appendix~\ref{app:errors}, were we extend the top and bottom panels of Fig.~\ref{fig:zz_agnostic}(c) for more values of $\gamma/\kappa$ and $C$, respectively, to unveil the different sources of errors in the total $\mathcal{I}$).

\subsection{XX interaction~\label{subsec:XXabstract}}

\subsubsection{Hamiltonian dynamics~\label{subsubsec:XXcoherent}}

\begin{figure}[tb]
    \centering
    \includegraphics[width=\linewidth]{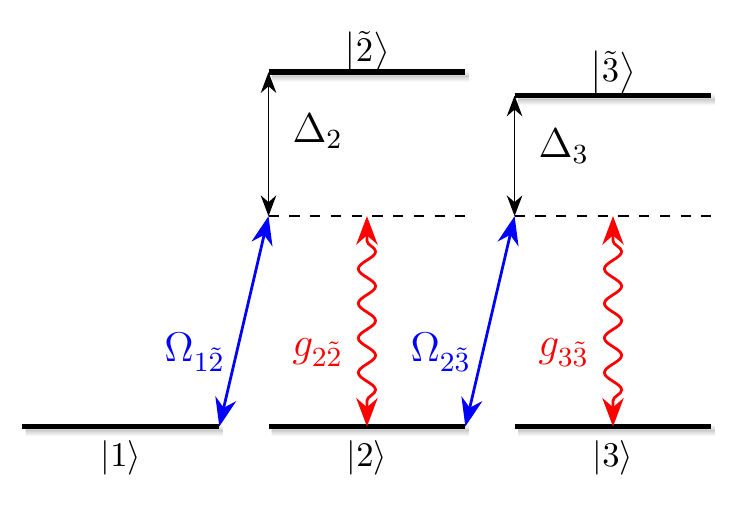}
\caption{Minimal multi-level configuration to obtain an effective spin 1 XX-type interaction between emitters coupled to a single mode photonic field. The transition $\ket{1}\leftrightarrow\ket{\tilde{1}}$ is not shown here since it does not enter in the effective dynamics.}
    \label{fig:int_XX_scheme}
\end{figure}

Let us now consider how to engineer one type of interaction which involves explicitly all the ground state levels, e.g., the XX Hamiltonian $S_x^i S_x^j+S_y^i S_y^j=\left(S^{i}\right)^\dagger S^j+\left(S^{j}\right)^\dagger S^i$, with $S_{x/y}^i$ being the spin-1 operator (once again, we will focus here on the purely Hamiltonian case and leave the discussion concerning the non-unitary dynamics to section \ref{subsubsec:XXlosses}). These operators can be codified in a ground-state manifold like the one depicted in Fig.~\ref{fig:int_XX_scheme} as follows:
\begin{eqnarray}
S^{j}_x &&= \frac{\s{1}{2}{j}+\s{2}{3}{j}+\text{H.c.}}{\sqrt{2}}\,,\nonumber\\ 
S^{j}_{y}&&=\frac{-i\s{1}{2}{j}-i\s{2}{3}{j}+\text{H.c.}}{\sqrt{2}}\,.
\end{eqnarray}

To obtain that type of interactions, one needs to induce photon-exchange processes which change the atomic state from $m\leftrightarrow m\pm 1$. Thus, the simplest configuration for the optically excited states which can do that is the one depicted in Fig.~\ref{fig:int_XX_scheme}, in which the Raman lasers connect the excited states through the following Hamiltonian terms:
\begin{equation}\label{eq:XX_laser_term}
    H_L = \sum_{i}\Bigg[\frac{\oms{2}{1}}{2}\s{1}{\tilde{2}}{i}+\frac{\oms{3}{2}}{2}\s{2}{\tilde{3}}{i}\Bigg]+\mathrm{H.c.}
\end{equation}
whereas the light-matter interaction term reads:
\begin{equation}\label{eq:XX_field_term}
    H_I=\sum_{i} 
    \g{2}^i\s{2}{\tilde{2}}{i}a^\dagger+\g{3}^i\s{3}{\tilde{3}}{i}a^\dagger+\mathrm{H.c.}\,,
\end{equation}
both already written in a rotating frame with the laser frequency $\omega_L$.  Note, we did not write explicitly the $1\leftrightarrow\tilde{1}$ transition because with that laser configuration it will not play a role in the dynamics. Moreover, the jump (or Linblad) operators considered in this case need to include the two possible channels of decay from the excited states to the ground ones via spontaneous emission, $L_{\gamma_2,1}=\sum_i\sqrt{\gamma_2}\s{1}{\tilde{2}}{i}$, $L_{\gamma_2,2}=\sum_i\sqrt{\gamma_2}\s{2}{\tilde{2}}{i}$, $L_{\gamma_3,1} = \sum_i \sqrt{\gamma_3}\s{2}{\tilde{3}}{i}$ and $L_{\gamma_3,2}=\sum_i \sqrt{\gamma_3}\s{3}{\tilde{3}}{i}$, but also need to include the loss of photons $L_\kappa = \sqrt{\kappa}a$.

Furthermore, the Hamiltonian of the fast subspace has the same form than the one written in Eq.~\eqref{eq:int_zz_ex}. Then, under the conditions in which the fast subspace levels can be adiabatically eliminated, $\mathrm{max}|\Omega_{\alpha\tilde{\beta}}|\ll \max\Delta_i$ for the atomic case ($i=1,2,3$) and $\mathrm{max}|g_{\alpha\tilde{\beta}}|\ll \mathrm{max}\left|\Delta_a\right|$ for the field case, one arrives to the following effective Hamiltonian for the ground state subspace (see Apppendix~\ref{app:XX} for details):
\begin{eqnarray}\label{eq:int_XX_eff}
    H_\mathrm{eff} &&= H_\text{s}  -\sum_{i}\Big(\frac{\Delta_2|\oms{2}{1}|^2}{4\Delta^2_2+4\gamma_2^2}\s{1}{1}{i}+\frac{\Delta_3|\oms{3}{2}|^2}{4\Delta^2_3+4\gamma_3^2}\s{2}{2}{i}\nonumber\\
    &&+\frac{4\Delta_a\left|\xi_i\right|^2}{4 \Delta_a^2+\kappa^2}\s{1}{1}{i}+\frac{4\Delta_a\left|\eta_i\right|^2}{4 \Delta_a^2+\kappa^2}\s{2}{2}{i}\Big)+H_\text{int},
\end{eqnarray}
where the interaction term $H_\text{int}$ now reads:
\begin{eqnarray}\label{eq:int_XX_int}
    H_\text{int} = -\frac{4\Delta_a}{4\Delta_a^2+\kappa^2}&&\sum_{i\neq j}\Big[\xi_i\bar{\xi_j}\s{2}{1}{i}\s{1}{2}{j}+\eta_i\bar{\eta_j} \s{3}{2}{i}\s{2}{3}{j}\nonumber\\
         &&+\xi_i\bar{\eta_j}\s{2}{1}{i}\s{2}{3}{j}+\bar{\xi_j}\eta_i\s{3}{2}{i}\s{3}{2}{j}\Big],
\end{eqnarray}
written in terms of the parameters
\begin{equation}\label{eq:xieta_def}
    \xi_i = \frac{2\bar{\Omega}_{1\tilde{2}}\g{2}^i\Delta_2}{4\Delta_2^2+4\gamma_2^2}\quad\text{and}\quad \eta_i = \frac{2\bar{\Omega}_{2\tilde{3}}\g{3}^i\Delta_3}{4\Delta_{3}^2+4\gamma_3^2}.
\end{equation}

Setting $\xi_i = \eta_i$, the interaction Hamiltonian then becomes the desired XX Hamiltonian:
\begin{equation}\label{eq:eff_xx_1st}
    H_\text{int} = \sum_{i< j}J_\text{xx}^{ij}\left(S^{i}_{x}S^{j}_x + S^{i}_y S^{j}_y\right)\,,
\end{equation}
with overall strength given by;
\begin{equation}\label{eq:eff_xx_1st_J}
    J_\text{xx}^{ij}=\Re\left[\frac{-4\xi_i\bar{\eta_j}\Delta_a}{4\Delta_a^2+\kappa^2}\right]\,.
\end{equation}

However, differently from the ZZ ones, where the energy shifts between the $1,3$ levels become equal for $\mu_i=-\nu_j$, here the Stark-shifts generated by the lasers (first line in Eq.~\eqref{eq:int_XX_eff}) are different for the condition $\xi_i = \eta_i$. This generates an effective detuning between the $2\leftrightarrow 1$ transition which in the limit where $\gamma_{i}\ll |\Delta_i|$ reads:
\begin{align}
    \delta_{21}\approx \frac{|\Omega_{2\tilde{3}}|^2}{4\Delta_3}\left(1-\frac{g_{3\tilde{3}}}{g_{2\tilde{2}}}\frac{\Delta_{2}}{\Delta_{3}}\right)\,.
\end{align}

Such detuning can then be cancelled by choosing $g_{3\tilde{3}}\Delta_2\approx g_{2\tilde{2}}\Delta_3$, or by connecting one of the two states ($2$ or $1$) to another far detuned excited state to generate another additional AC stark shift of the same order and opposite sign. However, in the different realizations, such as the ones discussed in section \ref{sec:atoms}), other AC Stark shifts than the ones discussed here could appear. Thus, the particular way to deal with these shifts would depend on the emitter chosen.

As a final remark, in the cases where the laser driving is linearly polarized and couple states $m\leftrightarrow \tilde{m}$, it should be the cavity/nanophotonic transition the ones that couple $m\leftrightarrow \tilde{m\pm 1}$. However, in that case one typically needs to include both the left/right circularly polarized modes, see e.g., Ref.~\cite{Orioli2021}, unless one exploits chiral quantum optic setups~\cite{lodahl17a}. In any case, the resulting interactions can also be made to emulate an effective XX Hamiltonian.

\subsubsection{Non-unitary contributions~\label{subsubsec:XXlosses}}

\begin{figure}
    \centering
   \includegraphics[width=\columnwidth]{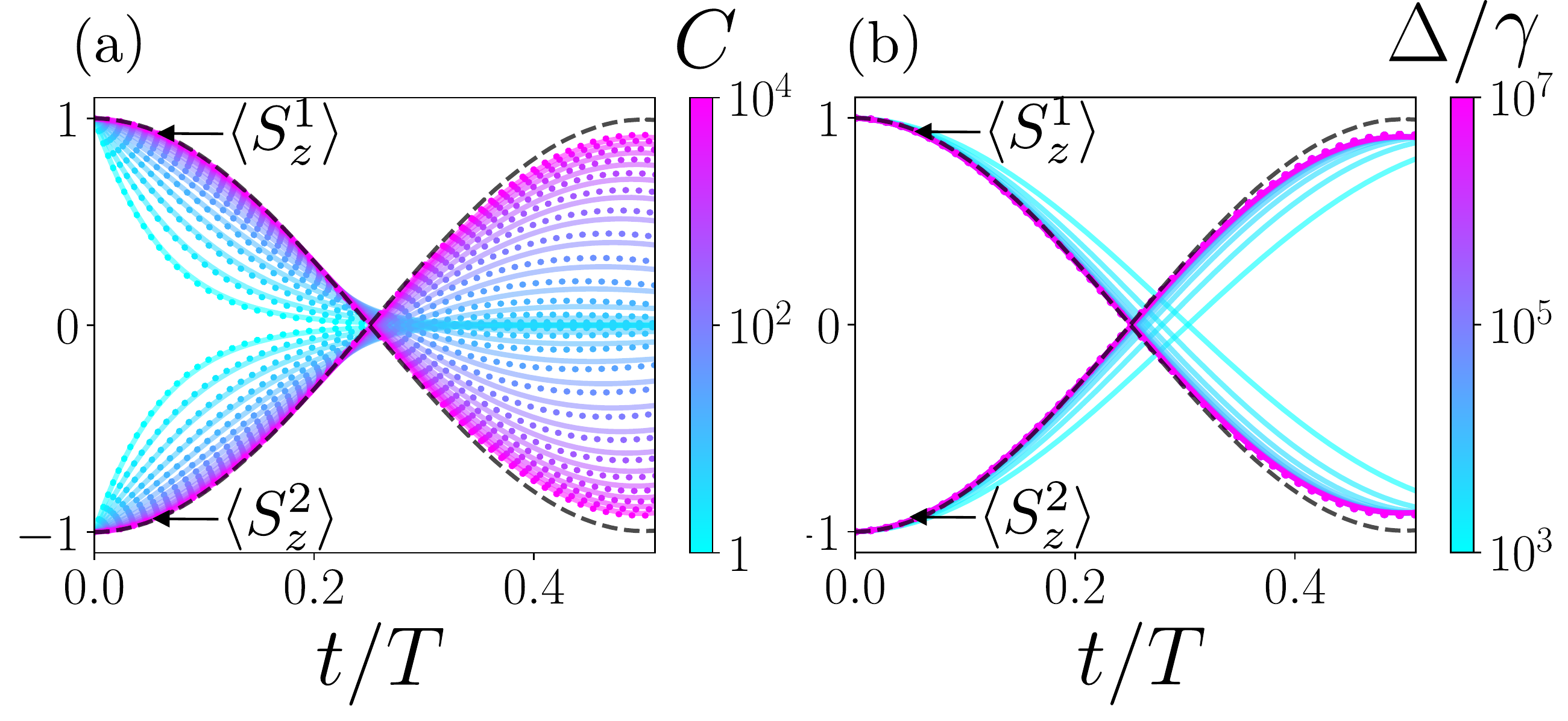}
    \caption{(a) Values of $\langle S_{x}^{i}\rangle$ as function of time (during half period, $T=2\sqrt{2}\pi/J$ for the considered initial state) for two multilevel quantum emitters interacting with a quantized mode and driven with an external field, according to the scheme in Fig.~\ref{fig:int_XX_scheme} and including losses, for different values of the cooperativity $C=g^2/(\kappa\gamma)$. The solid lines show a calculation considering the whole Hilbert space, while the dots correspond to the effective operators obtained in Eqs. \eqref{eq:eff_xx_1st} and \eqref{eq:losses_XX_1}-\eqref{eq:losses_XX_2} (for the coherent part and the losses, respectively, while removing the AC Stark shifts of the ground state manifold and taking $H_\text{s}=0$). The initial state was $\ket{\Psi(0)}=\ket{m=1}_{z}\otimes \ket{m=-1}_z$, and the parameters were $\Delta_2 = \Delta_3 := \Delta$, $\Omega_{1\tilde{2}} = \Omega_{2\tilde{3}} = \Delta/20$, $\Delta_a = -100\Delta$, $g_{2\tilde{2}}=g_{3\tilde{3}}=\sqrt{\gamma/\kappa}\Delta_a$ (so $\Delta_a$ minimizes the losses) and $\gamma$ and $\kappa$ were taken as described in figure \ref{fig:zz_agnostic}. (b) Same as in (a), but keeping the cooperativity constant at $C=10^{4}$ while modifying the rate of spontaneous emission, $\gamma/\Delta$ keeping $\kappa=|\Delta^{\text{opt}}_{a}|/\sqrt{C}$ to obtain the optimal detuning. Note that the effective evolutions collapse onto the same dots.\\
    In both figures we have plotted the ideal coherent evolution with a black dashed line.}
    \label{fig:agnostic_xx}
\end{figure}

Regarding the effective jump operators of the induced evolution, we find the operators 
\begin{eqnarray}\label{eq:losses_XX_1}
    L_{\gamma_j,1}^{\text{eff}}&&=\sum_i \frac{\Omega_{j-1,\tilde{j}}\sqrt{\gamma_j}}{2\Delta_j-2i\gamma_j}\sigma_{j-1,j}\quad\text{and}\\
    L_{\gamma_j,2}^{\text{eff}}&&=\sum_i \frac{\Omega_{j-1,\tilde{j}}\sqrt{\gamma_j}}{2\Delta_j-2i\gamma_j}\sigma_{j-1,j}
\end{eqnarray}
for the spontaneous decay from the atomic excited states, whereas the photon losses induce another one given by:
\begin{equation}\label{eq:losses_XX_2}
    L_{\kappa}^{\text{eff}}=\frac{-2\sqrt{\kappa}}{2\Delta_a - i\kappa}\sum_i \left(\xi_i \s{2}{1}{i}+\eta_i \s{3}{2}{i}\right).
\end{equation}

These operators are proportional to the same parameters than the ones derived on the previous section for the ZZ interaction, so an identical analysis concerning the losses as the one in subsection \ref{subsubsec:ZZlosses} could be made to find the same scaling of the expected errors $\epsilon_{\text{losses}}\sim1/\sqrt{C}$, again competing against the accuracy of the adiabatic elimination (as it can be seen in figures \ref{fig:agnostic_xx}a and \ref{fig:agnostic_xx}b). Moreover, note that for the high cooperativity case ($C=10^3-10^4$) and considering a moderate rate of spontaneous emission ($\gamma/\Delta\lesssim 10^{-5}$) both the evolutions considering the whole Hilbert space (with solid lines) and the effective ones (with dots) match the ideal coherent evolution (represented with black dashed lines), although this agreement is worse as time increases. This situation justifies a simplified analysis in section \ref{sec:XXatoms}, where we will study the capability of our system to entangle a pair of atoms considering an ideal coherent evolution, since one only requires the first quarter of the period ($t/T=1/4$, where $T=2\sqrt{2}\pi/J$ for this initial state).

\subsubsection{Recovering an Ising-type interaction}

\begin{figure}[tb]
    \centering
    \includegraphics[width=\linewidth]{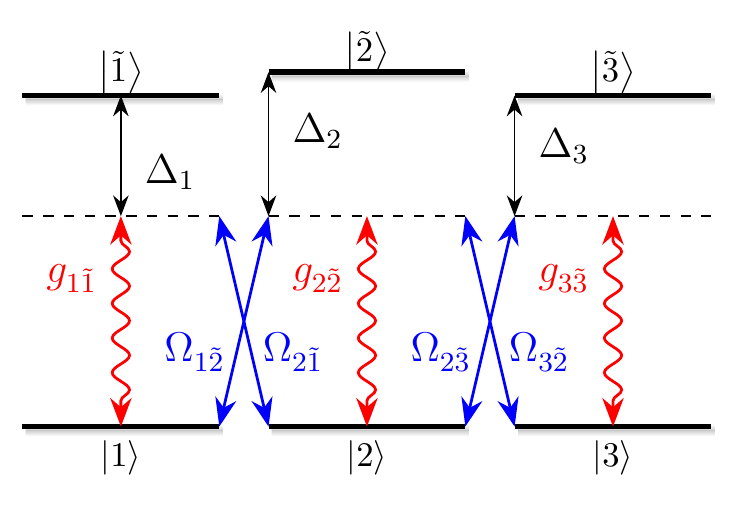}
\caption{Multi-level configuration to obtain an effective spin 1 Ising-type interaction between emitters coupled to a single mode photonic field, as a minimal variation of the setup in figure \ref{fig:agnostic_xx} used to obtain an XX interaction.}
    \label{fig:int_XXising_scheme}
\end{figure}

Furthermore, as a fundamental difference with the two-level case, the rich multilevel manifold that we have in this case allows the use of more ingredients in our models, leading to different ways to obtain the same interactions (which can be more or less useful depending on the experimental platform considered). In particular, we can slightly modify the configuration proposed in Fig.~\ref{fig:int_XX_scheme} including another Raman laser driving transitions in the opposite direction, as shown in Fig.~\ref{fig:int_XXising_scheme}. This modification differs from the previous one in two aspects: first, the transition $\ket{1}\leftrightarrow\ket{\tilde{1}}$ now enters in the dynamics; and, second, the driving term in Eq.~\eqref{eq:XX_laser_term} becomes
\begin{equation}
    H_L = \sum_{i}\frac{\oms{2}{1}}{2}\s{1}{\tilde{2}}{i}+\frac{\oms{3}{2}}{2}\s{2}{\tilde{3}}{i}+\frac{\oms{2}{1}}{2}\s{2}{\tilde{1}}{i}+\frac{\oms{2}{3}}{2}\s{3}{\tilde{2}}{i}+\mathrm{H.c.}
\end{equation}
The discussion is almost equal to the one in section \ref{subsubsec:XXcoherent} (for simplicity in the notation, now we only take into account the jump operators inducing $\pi$ transitions), where the effective Hamiltonian projected onto the slow/ground subspace reads now
\begin{eqnarray}\label{eq:int_XXising_eff}
    H_\mathrm{eff} &&= H_\text{s} -\sum_{i}\Big(\frac{\Delta_2|\oms{2}{1}|^2}{4\Delta^2_2+\gamma_2^2}\s{1}{1}{i}+\frac{\Delta_3|\oms{3}{2}|^2}{4\Delta^2_3+\gamma_3^2}\s{2}{2}{i}\\&&+\frac{\Delta_1|\oms{2}{1}|^2}{4\Delta^2_1+\gamma_1^2}\s{2}{2}{i}
    +\frac{\Delta_2|\oms{3}{2}|^2}{4\Delta^2_2+\gamma_1^2}\s{3}{3}{i}+\frac{4\Delta_a\left|\xi_i\right|^2}{4 \Delta_a^2+\kappa^2}\s{1}{1}{i}\nonumber\\&&+\frac{4\Delta_a\left(\left|\eta_i\right|^2+\left|\zeta_i\right|^2\right)}{4 \Delta_a^2+\kappa^2}\s{2}{2}{i}+\frac{4\Delta_a\left|\varphi_i\right|^2}{4 \Delta_a^2+\kappa^2}\s{3}{3}{i}\Big)+H_\text{int}\,,\nonumber
\end{eqnarray}
with an interaction term $H_\text{int}$ that reads:
\begin{eqnarray}\label{eq:int_XXising_int}
    H_\text{int} = -&&\frac{4\Delta_a}{4\Delta_a^2+\kappa^2}\sum_{i\neq j}\Big(\xi_i\s{2}{1}{i}+\eta_i \s{3}{2}{i}+\varphi_i \s{2}{3}{i}+\zeta_i \s{1}{2}{i}\Big)\nonumber\\
    &&\times\Big(\bar{\xi}_j\s{1}{2}{j}+\bar{\eta}_j\s{2}{3}{j}+\bar{\varphi}_j\s{3}{2}{j}+\bar{\zeta}_j\s{2}{1}{j}\Big).
\end{eqnarray}
The condition to be imposed now is that $\xi_i = \eta_i = \varphi_i = \zeta_i$, and this leads to an interaction term of the form:
\begin{equation}\label{eq:XXising_effH}
    H_\text{int}=\sum_{i<j} J_\text{Ising}^{ij}S_{x}^{i}S_{x}^{j},
\end{equation}
with strength
\begin{equation}
    J_\text{Ising} = \Re\left[\frac{-8\xi_i\bar{\xi}_j\Delta_a}{4\Delta_a^2+\kappa^2}\right].
\end{equation}
The Hamiltonian in Eq.~\eqref{eq:XXising_effH} is an Ising-type interaction, but now it is defined along the X axis (instead of the Z axis, as it was the case for the ZZ interaction derived in section \ref{subsec:ZZabstract}). On the one hand, this feature highlights the versatility of multilevel emitters to obtain different type of interactions. Besides, the similarities between the configurations shown in Figs.~\ref{fig:agnostic_xx} and \ref{fig:int_XXising_scheme}, that only differ by a single laser, allow for a simple transition from the XX interaction to the Ising one, which can be used to perform digital-analog quantum simulation of spin models with a reduced number of Trotter steps (see Section \ref{sec:applications} for more details).

\section{Particularizing to atomic emitters~\label{sec:atoms}}

After having derived the general conditions to obtain ZZ and XX spin-1 Hamiltonians, now we will particularize for specific multi-level quantum emitters. A particularly appealing system is that of Alkali atoms, such as Rubidium, where a natural ground and excited-state multi-level structure appears due to the coupling between the electronic and nuclear degrees of freedom. This coupling generates a set of multiplets in the ground (and optically excited) states characterized by their total angular momentum, $F$ ($\tilde{F}$), which are well-separated in energies, and thus, can be addressed independently. Besides, each hyperfine level contains $2F+1$ ($2\tilde{F}+1$) degenerate states distinguished by their angular momentum projection over a fixed axis $m_F (m_{\tilde{F}})$, which  we take as $\hat{z}$ without loss of generality, and which can be labelled as $m_F=-F,-F+1,\dots, F-1,F$. Note, these levels can also be separated in energy by applying a magnetic field. Thus, a ground state with hyperfine angular momentum $F=1$ represents an excellent candidate to encode the spin-1 system that we require to obtain the interactions developed in the previous Section~\ref{sec:general}.

However, a complication arises from the angular momentum origin of these subspaces, that are, the connection between the ground and excited levels have certain limitations imposed by the optical selection rules and Clebsch-Gordan (CG) coefficients. For example, optical selection rules only allow transitions such as $m_{\tilde{F}}=m_F+q$, with $q\in\{0,\pm 1\}$ being the units of angular momentum that the photon mediating the transition carries. Besides, each transition is weighted by a different CG coefficient given by: $C_{q}^{m_F} \equiv \braket{F,m_F;1,q|\tilde{F},m_F+q}$. This means that, if one shines a laser with a frequency $\omega_L$ and a given amplitude over the atoms, the driving term must be written as:
\begin{equation}\label{eq:hyperfine_theory_driving}
    H_{\text{d}}^{q} = \sum_i\left( \frac{\Omega}{2}S^{i\,\dagger}_q e^{-i\omega_L t}+\mathrm{H.c.}\right)\,,
\end{equation}
where:
\begin{equation}
    S^{i\,\dagger}_{q} = \sum_{m_g} C_{q}^{m_F} \sigma_{\tilde{F} m_F -q,F m_F}^{i}\,,
\end{equation}
being $\sigma_{\tilde{F} m_q -q,F m_q}^{i}=\ket{\tilde{F},m_F - q}_i\bra{F,m_F}$ the atomic coherence operator between states $\ket{F,m_F}$ and $\ket{\tilde{F},m_F-q}$. The same happens with the light-matter interaction Hamiltonian which has also to be weighted by the same coefficients:
\begin{eqnarray}\label{eq:hyperfine_theory_coupling}
    H_I^q = g\sum_{i} \left(S^{i\,\dagger}_q a_q+\mathrm{H.c.}\right)
\end{eqnarray}
(in the following, we will include the Clebsch-Gordan coefficients into the amplitudes for each atomic transition). Overall, these limitations make that generating the simplified level structure depicted in Figs.~\ref{fig:int_zz_scheme},\ref{fig:int_XX_scheme} requires additional engineering. In what follows, we explain how to obtain it using the Rb or Na atomic level structure.

\subsection{Engineering ZZ interactions~\label{subsec:ZZatoms}}

\begin{table}[tb]
    \centering
    \begin{tabular}{|c||c|c|c|c}
    \hline
             & $m_F = 1$ & $m_F = 0$ & $m_F = -1$\\\hline
        $\tilde{F}=2$ & $\sqrt{1/4}$ & $\sqrt{1/3}$ & $\sqrt{1/4}$ \\\hline
        $\tilde{F}=1$ & $-\sqrt{1/12}$ & $0$ & $\sqrt{1/12}$ \\\hline
    \end{tabular}
    \caption{Clebsch-Gordan coefficients corresponding to the $|\tilde{F},m_{\tilde{F}}\rangle\rightarrow |F=1,m_F\rangle$ $\pi$ transitions in the $D_1$ line of ${}^{87}\text{Rb}$ taken from Ref.~\cite{Steck2001}.}
    \label{tab:CG_pi}
\end{table}

\begin{figure}[tb]
    \includegraphics[width=\linewidth]{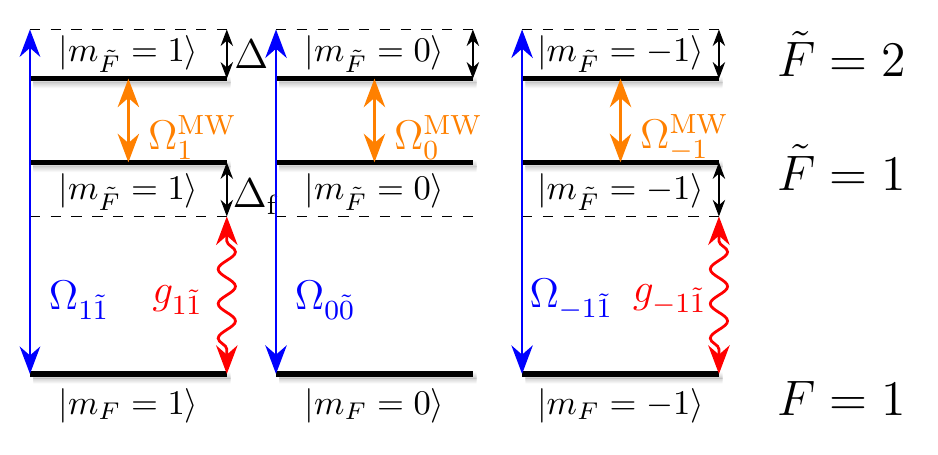}
    \caption{Proposed configuration to obtain an effective ZZ interaction between multilevel quantum emitters. It uses the $D_1$ line in ${}^{87}\text{Rb}$. The excited levels with $m_{\tilde{F}}=\pm 2$ have not been plotted here since they do not play a role in the proposed scheme.}
    \label{fig:imp_zz}
\end{figure}

\begin{table}[tb]
    \centering
    \begin{tabular}{|c||c|c|c|c}
    \hline
             & $m_{\tilde{F}} = 1$ & $m_{\tilde{F}} = 0$ & $m_{\tilde{F}}= -1$\\\hline
        $\tilde{F}=2$ & $-\sqrt{3/4}$ & $-1$ & $-\sqrt{3/4}$ \\\hline
    \end{tabular}
    \caption{Clebsch-Gordan coefficients corresponding to the $|\tilde{F},m_{\tilde{F}}\rangle\rightarrow |\tilde{F}=1,m_F\rangle$ $\pi$ microwave transitions in the $D_1$ line of ${}^{87}\text{Rb}$ taken from Ref.~\cite{Boguslawski2019}.}
    \label{tab:MW_pi}
\end{table}

Let us start with the simpler case of the ZZ interactions. According to the level scheme of Fig.~\ref{fig:int_zz_scheme}, one requires that linearly polarized lasers which connect the $m_F=\pm 1 \leftrightarrow m_{\tilde{F}}=\pm 1 $ transitions, while leaving the state $m_F=0$ unaltered. Interestingly, according to the selection rules (see Table~\ref{tab:CG_pi}) the transition $\ket{F=1,m_F=0}\leftarrow \ket{\tilde{F}=1,m_{\tilde{F}}=0}$ is forbidden, so that condition would come for free by considering a $F=1\leftarrow \tilde{F}=1$ level scheme. However, the other condition is that $\mu_i=-\nu_i$ in Eq.~\eqref{eq:munu_def} so that the system engineers a perfect ZZ interaction. Since the CG coefficient between the $m_F=\pm 1 \leftrightarrow m_{\tilde{F}}=\pm 1 $ transitions that appears in both $g_{ii}$ and $\Omega_{ii}$ is equal in absolute value, the only way to achieve that condition with only that levels would be imposing $\Delta_{-1}=-\Delta_{1}$. In principle, one can achieve that by using a magnetic field that yields a Zeeman splitting between the $m_{\tilde{F}}=\pm 1$ at a rate $-0.2$ MHz/G. However, this also breaks the degeneracy in the ground state subspace at even larger rate $-0.7$ MHz/G, which breaks the assumptions under which the effective dynamics of Eqs.~\eqref{eq:theory_master_eff}-\eqref{eq:theory_effH} is obtained. Thus, one needs to search for alternatives.

\begin{figure}[tb]
    \centering
    \includegraphics[width=\columnwidth]{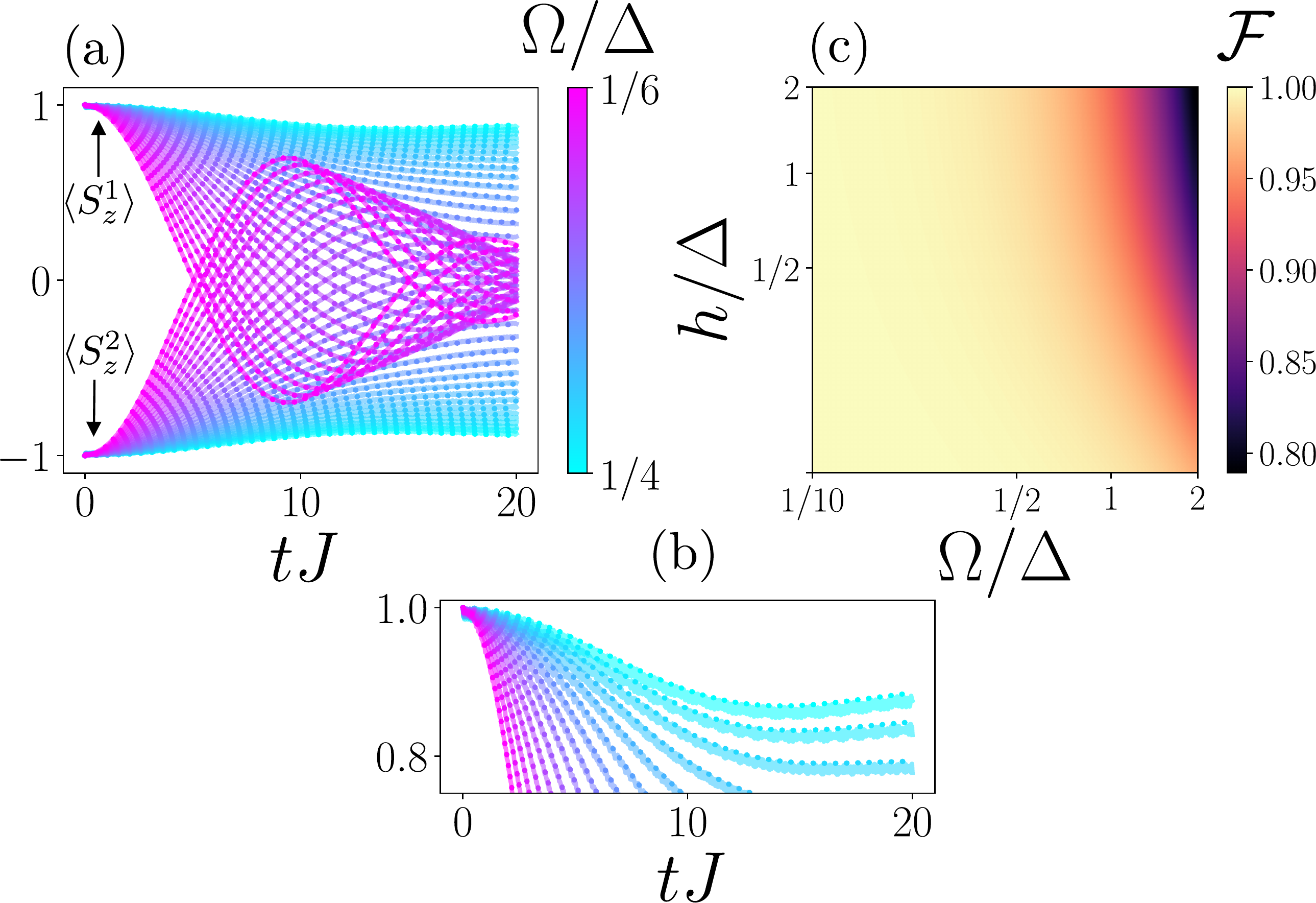}
    \caption{(a) Mean values of the $\left\langle S^{i}_{z}\right\rangle$ operators for two atoms interacting with the electromagnetic field as a function of time for different driving strenghts $\Omega$ (equation \eqref{eq:hyperfine_theory_driving}). The solid lines are the numerical results obtained with a full calculation that considers both the slow and the fast subspaces whereas the dots correspond to the effective evolution given by \eqref{eq:eff_ising}, taking in both cases $H_\text{s} = h\sum_i S^{i}_x $. The initial state was $\ket{\Psi(0)}=\ket{m=1}\otimes\ket{m=-1}$, while the parameters were $\Delta_1 = \Delta_2 :\equiv \Delta$ (the microwave in resonance with the $\tilde{F}=2\leftrightarrow\tilde{F}=1$ transition), $\Delta_\text{f}/\Delta =2$, $\Omega_\text{MW}=-2\Omega$, $g/\Delta = 1/5$ and $h/\Delta = 5/3\times 10^{-7}$. The amplitude $J$ of the effective Ising evolution for each case was of order $J/\Delta\sim 10^{-7}$ (given by equation \eqref{eq:eff_ising_Jz}. (b) Close detail of panel (a) showing how the accuracy of the effective evolution decreases as $\Omega/\Delta$ increases (we show here half of the lines). (c) Fidelity, defined as $\mathcal{F}=\mathrm{Tr}\left(\sqrt{\sqrt{\rho}\rho_\mathrm{eff}\sqrt{\rho}}\right)^2$, between the ground state of a 2-qutrits spin-1 Ising model according to equation \eqref{eq:eff_ising} and the ground state of the full Hamiltonian including also the excited subspaces and the cavity mode for two atoms (tracing the extra degrees of freedom). These simulations have been made setting $\kappa=\gamma_i=0$.}
    \label{fig:int_zz}
\end{figure}

One possibility is depicted in Fig.~\ref{fig:imp_zz}: it consists in using the $F=1\leftrightarrow \tilde{F}=1$ to couple to the cavity/nanophotonic mode to take advantage of the forbidden transition, but then driving it through an effective two-photon transition through the $F=2$ manifold, i.e.,  $\ket{F=1,m_F=\pm 1}\leftrightarrow \ket{\tilde{F}=2,m_{\tilde{F}}=\pm 1}\leftrightarrow \ket{\tilde{F}=1,m_{\tilde{F}}=\pm 1}$, using a laser (microwave) field $\Omega_{m_F m_{\tilde{F}}}$ $\left(\Omega_{m_F}^\mathrm{MW}\right)$, respectively. Under the conditions that the $\tilde{F}=2$ states can be adiabatically eliminated (see Appendix~\ref{app:derivations}), the resulting effective driving between the $F=1\leftarrow \tilde{F}=1$ transition reads:
\begin{equation}
    H^\text{eff}_L =\sum_{i}\left(\frac{\Omega^{\text{eff}}_{1}}{2}\s{1}{\bar{1}}{i}+\frac{\Omega^{\text{eff}}_{2}}{2}\s{2}{\bar{2}}{i}+\frac{\Omega^{\text{eff}}_{3}}{2}\s{3}{\bar{3}}{i}+\mathrm{H.c.}\right),
\end{equation}
with $\Omega^{\text{eff}}_{j}\equiv -2\Omega_{j\bar{j}}\Omega^\text{MW}_{j}\Delta_2/(4\Delta_{2}^{2}+\gamma_{2}^2)$, with $\gamma_2$ the rate of spontaneous emission from that level and were we have introduced the detunings from the $\tilde{F}=1$ and $\tilde{F}=2$ transitions, respectively, as
\begin{equation}
    \Delta_1 = \omega_1 - \omega_L + \omega_{MW}\quad\text{and}\quad\Delta_2 = \omega_2 - \omega_L\,,    
\end{equation}
were we also include the effect of the microwave driving (the details can be found in Appendix~\ref{app:ZZ_at}). The key point is that the CG coefficients for the laser transition $\ket{F=1,m_F=\pm 1}\leftrightarrow \ket{F=2,m_F=\pm 1}$ and microwave transitions are equal (see Tables~\ref{tab:CG_pi}-\ref{tab:MW_pi}), whereas the ones of the transition $\ket{F=1,m_F=\pm 1}\leftrightarrow \ket{F=1,m_F=\pm 1}$ have different sign. This makes that $\Omega_{1}^\mathrm{eff}=\Omega_{-1}^\mathrm{eff}$ while $g^i_{11}=-g^i_{33}$, the parameters required to satisfy the condition $\mu_i=-\nu_i$ of Eq.~\eqref{eq:munu_def}. Note also that with that effective driving $H^\text{eff}_L$ term, the Hamiltonian describing the physics are formally equivalent to the ones used in Section~\ref{subsec:ZZabstract}, except for the additional driving of $\ket{F=1,m_F=0}$ level because the transition to the $\ket{\tilde{F}=2,m_F=0}$ is not forbidden. Thus, by adiabatically eliminating the $\tilde{F}=1$ states and the photonic modes, we arrive to the desired ZZ Hamiltonian with an additional AC Stark-shift:
\begin{equation}\label{eq:eff_isingatom}
   H_\text{ZZ,eff} = H_\text{s}+ \sum_{i<j} J_z^{ij} S^{i}_z S^{j}_z+\sum_i \left(\delta_1-\delta_0\right) S_{z}^{i\,2}\,\,,
\end{equation}
with $J_z^{ij}$ being:
\begin{equation}
     \label{eq:Jijzz}
    J_z^{ij}=-\frac{2\Delta_a}{4\Delta_{a}^2+\kappa^2}\frac{\left|\Omega\right|^2\left|\Omega^{\text{MW}}\right|^2\left|g\right|^2\Delta_{2}^2\Delta_{1,\pm 1}^{\text{eff}\,2}}{\big(4\Delta_{2}^2+\gamma_{2}^{2}\big)^2\left(4\Delta_{1,\pm 1}^{\text{eff}\,2}+\gamma_{1}^{2}\right)^2}\,,
 \end{equation}
and
\begin{equation}
     \delta_k=\frac{4\left|\Omega_{{k}\tilde{k}}\right|^2 \left|\Omega_{k}^{\text{MW}}\right|^2\Delta_{2}^{2}\Delta_{1,k}^{\text{eff}}}{\big(4\Delta_{2}^2+\gamma_{2}^2\big)^2\big(4\Delta_{1,k}^{\text{eff}\,2}+\gamma_{1}^2\big)}\,,
\end{equation}
where $\Delta_{1,j}^{\text{eff}}=\Delta_1 - |\Omega_{j}^{\text{MW}}|^2\Delta_2/(4\Delta_{2}^{2}+\gamma^2)$ and $\Delta_a = \omega_a - \omega_L + \omega_{MW}$. Note that this AC stark-shift can be compensated by connecting off-resonantly with another excited state in such a way that generates another Stark-shift with the opposite sign.

Finally, let us benchmark numerically that indeed all the approximations we made are correct, such that the dynamics of the multi-level emitter of Fig.~\ref{fig:imp_zz} is captured by Hamiltonian $H_\mathrm{eff}$ of Eq.~\eqref{eq:eff_isingatom}. For that, we assume to be in the conditions where $\delta_1=\delta_0$, and consider that $H_\text{s}$ describes an additional transverse field of the spin-1 system, i.e., $H_\text{s}=h\sum_i S_x^i$, which can be obtained through a microwave field or with additional two-photon Raman-assisted processes. In Fig.~\ref{fig:int_zz}(a), we initialize two atoms in the $\ket{\psi(0)} = \ket{m=1}\otimes\ket{m=-1}$ state and let them evolve with the effective Hamiltonian of Eq.\eqref{eq:eff_isingatom} (dotted lines) and the full Hamiltonian of the multi-level structure of Fig.~\ref{fig:imp_zz} (solid lines) for a transverse field $h/J\sim 0.5$. The different colors indicate different driving amplitudes $\Omega/\Delta$ to show how the agreement becomes  better as $\Omega/\Delta\rightarrow 0$. Beyond the dynamics, another magnitude of interest that one can look is how well the Hamiltonian can capture the ground-states of the interacting models. This is what we illustrate in Fig.~\ref{fig:int_zz}(b), where we plot the fidelity of the ground state of the effective model vs the one of the full Hamiltonian. As expected again, the smaller $\Omega/\Delta$, the better is the agreement, since this is the regime where the adiabatic elimination is expected to work. We want to note that these simulations have been made assuming $\kappa=\gamma_i=0$ to focus on the effects on the deviations in the purely Hamiltonian simulation.

\subsection{Engineering XX interactions~\label{sec:XXatoms}}

\begin{figure}[tb]
    \includegraphics[width=\linewidth]{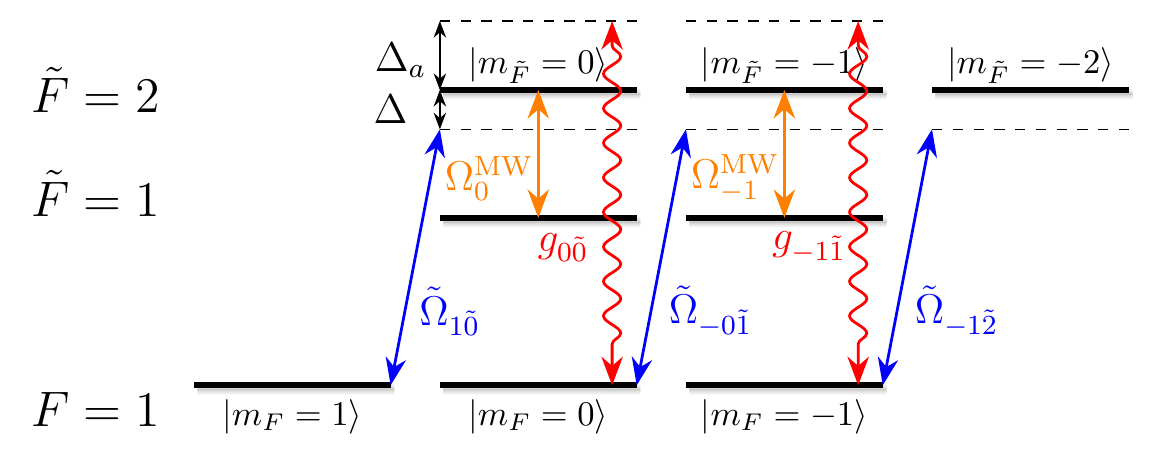}
    \caption{Proposed configuration to obtain an effective XX interaction between multilevel quantum emitters. It uses the $D_1$ line in ${}^{87}\text{Rb}$, and we set the microwave in resonance with the transition $\tilde{F}=1\leftrightarrow\tilde{F}=2$, so $\omega_{MW} = \omega_2 - \omega_1$ and hence $\Delta_1 = \omega_1-\omega_L+\omega_{MW}=\Delta_2 = \omega_2 - \omega_L \equiv\Delta$. The transitions involving the $m_{\tilde{F}}=1$ levels are not shown here since they do not enter into the effective dynamics.}
    \label{fig:imp_xx}
\end{figure}

To engineer the XX Hamiltonian we also use the multi-level structure appearing in the D1 line of Rubidium as depicted in Fig.~\ref{fig:imp_xx}. Differently from the ZZ case, now the cavity field should be closer to the $\tilde{F}=2$ levels, so that they couple preferentially to them. The reason is that in this case we need the transition between the $m_F=0\leftrightarrow m_{\tilde{F}}=0$ to be active, something that can not be obtained using the $\tilde{F}=1$-states as the intermediate states due to the optical selection rules. Besides, the laser connecting the ground ($F=1$) and optically excited states ($\tilde{F}=2$) must be circularly polarized, so that it enables the effective transitions $m_F\leftarrow m_F\pm 1$ exchanging a cavity/nanophotonic photon.  However, similarly to what occurs for the ZZ interaction, the different CG coefficients for the circularly polarized transitions (see Table~\ref{tab:CG_sigma-}), again complicates obtaining the desired conditions to achieve the pure XX Hamiltonian, i.e., $\xi=\eta$ in Eq.~\eqref{eq:xieta_def}. In fact, if the detuning between the three transitions is the same, achieving $\xi=\eta$ is not possible due to the different CGs of the $\ket{F=1,m_F=1 (0)}\leftrightarrow \ket{\tilde{F}=2,m_F=0 (-1)} \leftrightarrow \ket{F=1,m_F=0 (-1)}$ transitions.

\begin{table}[tb]
    \centering
    \begin{tabular}{|c||c|c|c|c}
    \hline
             & $m_F = 1$ & $m_F = 0$ & $m_F = -1$\\\hline
        $\tilde{F}=2$ & $-\sqrt{1/12}$ & $-\sqrt{1/4}$ & $-\sqrt{1/2}$ \\\hline
    \end{tabular}
    \caption{Clebsch-Gordan coefficients corresponding to the $|\tilde{F},m_{\tilde{F}}\rangle\rightarrow |F=1,m_F-1\rangle$ $\sigma^-$ transitions in the $D_1$ line of ${}^{87}\text{Rb}$ used in Fig.~\ref{fig:imp_xx}.}
    \label{tab:CG_sigma-}
\end{table}

A possible way out is to harness the $\tilde{F}=1$ levels as depicted in Fig.~\ref{fig:imp_xx} to induce a state-dependent AC Stark-shift on the excited levels via a resonant microwave field. In fact, if the intermediate level (detuned by $\Delta_1 \equiv \omega_1 - \omega_L + \omega_{MW}$) is adiabatically eliminated first, the effect over the the excited state levels of $\tilde{F}=2$ is to renormalize their energies by
\begin{align}
   \delta_{m_{\tilde{F}}}=-\frac{|C_{m_{\tilde{F}}}|^2|\Omega^\mathrm{MW}|^2}{4\Delta_1}\,,
\end{align}
where $C_{m_{\tilde{F}}}$ are the CG coefficients written in Table~\ref{tab:MW_pi}. Such state-dependent shifts add up to the effective laser detuning between the $F=1\leftarrow \tilde{F}=2$ transitions, $\Delta_2 \equiv \omega_2 - \omega_L$, and thus can be used a tuning knob to compensate the CG and enforce the condition $\xi=\eta$. In fact, in Appendix~\ref{app:XX_at} we show that the microwave amplitude for which this occurs is such that:
\begin{equation}\label{eq:imp_xx_cond}
    \left|\Omega_\text{MW}\right| = \sqrt{\frac{8}{3}\Delta_1 \Delta_2}.
\end{equation}

Imposing that value, one indeed obtains the desired XX interaction, plus a correction introduced by additional state-dependent AC stark-shifts that appear between the $F=2$ states:
\begin{align}\label{eq:effXX}
    H_{\mathrm{XX},\mathrm{eff}}=H_g+H_\text{Stark}+\sum_{i< j}J_\text{xx}^{ij}\left(S^{i}_{x}S^{j}_x + S^{i}_y S^{j}_y\right)\,,
\end{align}
where now $J_{\text{xx}}^{ij}$ reads (up to first order in $\gamma/\Delta$ and $\kappa/\Delta_a$)
\begin{equation}\label{eq:Jxx}
    J_{\text{xx}}^{ij}= -\frac{1}{16}\frac{\left|\Omega\right|^2 \left|g\right|^2}{\Delta^2\Delta_{a}}\,,
\end{equation}
where condition \eqref{eq:imp_xx_cond} is already satisfied and $\Delta_a = \omega_a - \omega_L + \omega_{MW}$. Furthermore, we have included the AC Stark shifts into the term $H_\text{Stark}$, that can be rewritten as:
\begin{equation}\label{eq:starkXX}
    H_\text{Stark} = \sum_i \epsilon_0 \mathbb{1}^i+\frac{\epsilon_1 - \epsilon_{-1}}{2}S_{z}^{i} + \frac{\epsilon_1 + \epsilon_{-1}-2\epsilon_0}{2}S_{z}^{i\,2}\,,
\end{equation}
where $\mathbb{1}_i$ is the unit operator and $\epsilon_i =\left|C_{i}\right|^2 \left|\Omega\right|^2/(4\Delta^\text{eff}_{2,i})$, with $C_i$ each of the  Clebsch-Gordan coefficients in Table \ref{tab:CG_sigma-}. These are different for each transition, so the Hamiltonian $H_\text{Stark}$ in Eq.\eqref{eq:starkXX} includes the terms proportional to $S_{z}^{i}$ and the terms proportional to $S_{z}^{i\,2}$, so both of them mask the effective evolution given by the pure XX Hamiltonian. In principle, two extra off-resonant lasers could be used to remove these terms, properly adjusting the detunings and amplitudes, but this introduces an extra experimental complication in our proposal. Thus, we consider a situation in which the AC Stark shifts are not completely removed, and instead only the condition $\epsilon_1 = \epsilon_{-1}$ is satisfied. This could be achieved using a  circularly polarized laser coupled off-resonantly to transitions in the $D_2$ line or even using a constant magnetic field to cancel the $S_{z}^{i}$ term using a Zeemann term appearing in $H_g$ (however, our conclusions regarding the effect of the remaining $\propto S_{z}^{i\,2}$ term can be easily applied to a case were the $\propto S_{z}^{i}$ term is not fully eliminated).
\begin{figure*}[t]
    \centering
   \includegraphics[width=\linewidth]{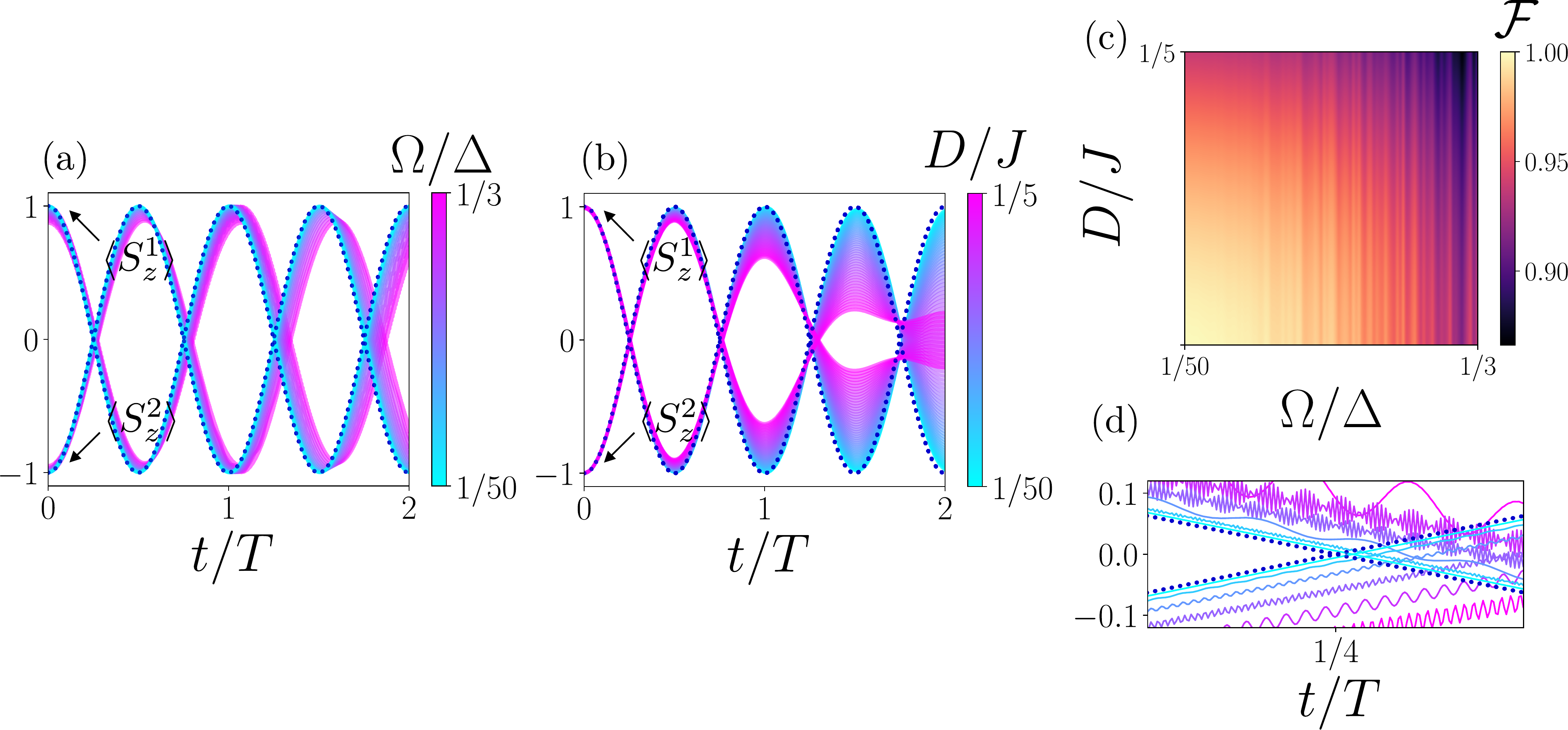}
    \caption{(a) Mean values of $\left\langle S^{i}_{x}\right\rangle$ for two atoms interacting with the electromagnetic field as a function of time for different driving strenghts $\Omega$ (Eq.~\eqref{eq:hyperfine_theory_driving}), where we have normalized the time using the period of coherent ideal oscillations for this initial state evolving only under $H_\text{int}$, $T=2\sqrt{2}\pi/J$, with $J:=J_\text{xx}^{ij}$, equal for all atoms $i$ and $j$, according to Eq.~\eqref{eq:eff_xx_1st_J}. Solid lines show numerical results obtained with a full calculation that considers both the slow and the fast subspaces and the dots correspond to the effective evolution given by Eq.~\eqref{eq:effXX}, taking in both cases $H_\text{s} = H_\text{Stark}=0$. The initial state was $\ket{\Psi(0)}=\ket{m=1}\otimes\ket{m=-1}$, while the parameters were $\Delta_1 = \Delta_2 :\equiv \Delta$ (the microwave in resonance with the $\tilde{F}=2\leftrightarrow\tilde{F}=1$ transition), $\Delta_\text{a}/\Delta =-2$, $\Omega_\text{MW}/\Delta=3\sqrt{8}$ (Eq.~\eqref{eq:imp_xx_cond}) and $g/\Delta = 1/15$. (b) Same as in (a), but fixing $\Omega/\Delta=1/10$ and instead modifying the value of the anisotropy introduced by a non-zero $H_\text{Stark}=D\sum_{i} S_{i}^{z\,2}$ Hamiltonian. (c) Fidelity, defined as $\mathcal{F}=\mathrm{Tr}\left(\sqrt{\sqrt{\rho}\rho_\mathrm{eff}\sqrt{\rho}}\right)^2$, between the state at $t=T/4$ of two spin-1 qutrits initially at $\ket{m=1}\otimes\ket{m=-1}$ evolving under the XX Hamiltonian without anisotropy (that is, only the term proportional to $J_{\text{xx}}^{ij}$ in Eq.~\eqref{eq:eff_ising}) and the state at $t=T/4$ of the atoms evoled under the full Hamiltonian, including the excited subspaces, the cavity and also a non-zero anisotropy $H_\text{Stark}$ proportional to $D$. (d) Inset of panel (a) at $t=T/4$ for only six of the amplitudes, showing how the small oscillations around the effective evolution increase as the amplitude $\Omega/\Delta$ grows. \\
    These simulations have been made setting $\kappa=\gamma_i=0$.}
    \label{fig:xx_atomic}
\end{figure*}

In the following, we numerically benchmark that indeed the dynamics of the effective model in Eq.~\eqref{eq:effXX} captures the physics of the full Hamiltonian and how it is able to entangle two atoms despite an incomplete elimination of the AC Stark shifts, which would lead to a term $H_\text{Stark}=D\sum_i S_{z}^{i\,2}$. To do so, we first start from an initial atomic state $\ket{\Psi(0)}=\ket{m_F=1}\otimes\ket{m_F=-1}$ and let it evolve under the full Hamiltonian (solid lines), showing also the effective evolution according only to the corresponding effective XX Hamiltonian, with $J$ given by Eq.~\ref{eq:Jxx} (with blue dots). Thus, in Fig.~\ref{fig:xx_atomic}(a) we first check the accuracy of the adiabatic elimination as the amplitude of the driving term is increased studying the values of $\langle S_{z}^{i}\rangle$. Remarkably, although a high-amplitude driving populates the excited levels and reduces the accuracy of the effective Hamiltonians, we find a very similar evolution up to the first quarter of period (that is only kept in the lower-amplitude case as time increases). Furthermore, this good agreement for shorter times is also kept when the AC Stark shifts are not fully eliminated, as it can be seen in Fig.~\ref{fig:xx_atomic}b.

Moreover, since we are interested in the generation of entangled atomic states, we also check the fidelity between the state resulting from the evolution under the full Hamiltonian and the one obtained with only the effective XX term at $t=T/4$ in Fig.~\ref{fig:xx_atomic}c, considering different driving amplitudes $\Omega/\Delta$ and also different anisotropies $D/J$. We find that small values of $\Omega/\Delta$ lead to fidelities up to $>95$\% even if the AC Stark shifts are not fully compensated. Finally, note that, for fixed $D/J$, the fidelity as a function of $\Omega/\Delta$ in Fig.~\ref{fig:xx_atomic}c shows oscillations when the amplitude increases. To understand this effect, we have included an inset of Fig.~\ref{fig:xx_atomic}a in Fig.~\ref{fig:xx_atomic}d, where it can be seen the smaller oscillations in $\langle S_{z}^{i}\rangle$ as $\Omega/\Delta$ grows, due to the increased population in the fast subspace. This leads to crossings in the observables, making some fidelities at $t=T/4$ higher even if the accuracy of the global adiabatic elimination would be worse.

\subsection{Differences between the cavity and nanophotonic setups~\label{subsec:nano}}

Although at the beginning of the manuscript we highlighted that our results could be applied to both atoms coupled to nanophotonic structures (Fig~\ref{fig:1}b) and inside a cavity (Fig~\ref{fig:1}c), we focused on the later for the derivations. The difference in the nanophotonic setup is that atoms can couple to any of the possible modes of a band with energy dispersion $\omega(\kk)$, and this modifies the final interaction amplitude and also its range. Now, we show the main differences between both setups, following closely Refs.~\cite{douglas15a,Gonzalez-Tudela2015b,Hung2016}. 

In particular, any product of terms of the form $g^i g^j\Delta_a/(4\Delta_{a}^2+\kappa^2)$ derived previously would need to be replaced by 
\begin{equation}
    f(\rr_i-\rr_j)=\frac{g^i g^j\Delta_a}{4\Delta_{a}^2+\kappa^2}\rightarrow\sum_{\kk\in\mathrm{1BZ}}\frac{\left|g_{\kk}\right|^2 \Delta_\mathbf{k}}{4\Delta_{\kk}^{2}+\kappa^2}e^{i\kk\cdot\left(\rr_i-\rr_j\right)},
\end{equation}
with $\Delta_\kk = \omega_\kk - \omega_L$ and $g_\kk$ the coupling constant between the atoms and the $\kk-$mode, and we have introduced the function $f(\rr_i-\rr_j)$ to characterize the coupling between atoms $i$ and $j$. Furthermore, if we assume the relevant atomic transitions are quadratic $\omega(\kk)$ (and also isotropic in the 2D case), we can get analytical expressions of the effective interactions by turning the sums into integrals. In particular, when the atomic frequency lies in the bandgap, the effective interactions between the emitters are mediated by an atom-photon bound state that leads to
\begin{equation}
    \left|f(\rr_i-\rr_j)\right| \propto e^{-\left|\rr_i-\rr_j\right|/L}
\end{equation}
for the case of a 1D waveguide and 
\begin{equation}
    \left|f(\rr_i-\rr_j)\right| \propto e^{-\left|\rr_i-\rr_j\right|/L}/\sqrt{\left|\rr_i-\rr_j\right|/L}
\end{equation}
when considering a 2D photonic crystal. Importantly, in both cases the effective interactions have a finite range $L$ (in contrast with the cavity QED setup, in which $|f(\rr_i-\rr_j)|=1$). However, this length, which corresponds to the atom-photon bound-state shape can be dynamically tuned adjusting system's parameters (such as the Raman laser frequency $\omega_L$). For more details we refer to the reader to Refs.~\cite{douglas15a,Gonzalez-Tudela2015b,Hung2016}.

\section{Potential applications~\label{sec:applications}}

    \begin{figure}[tb]
        \centering
        \includegraphics[width=\linewidth]{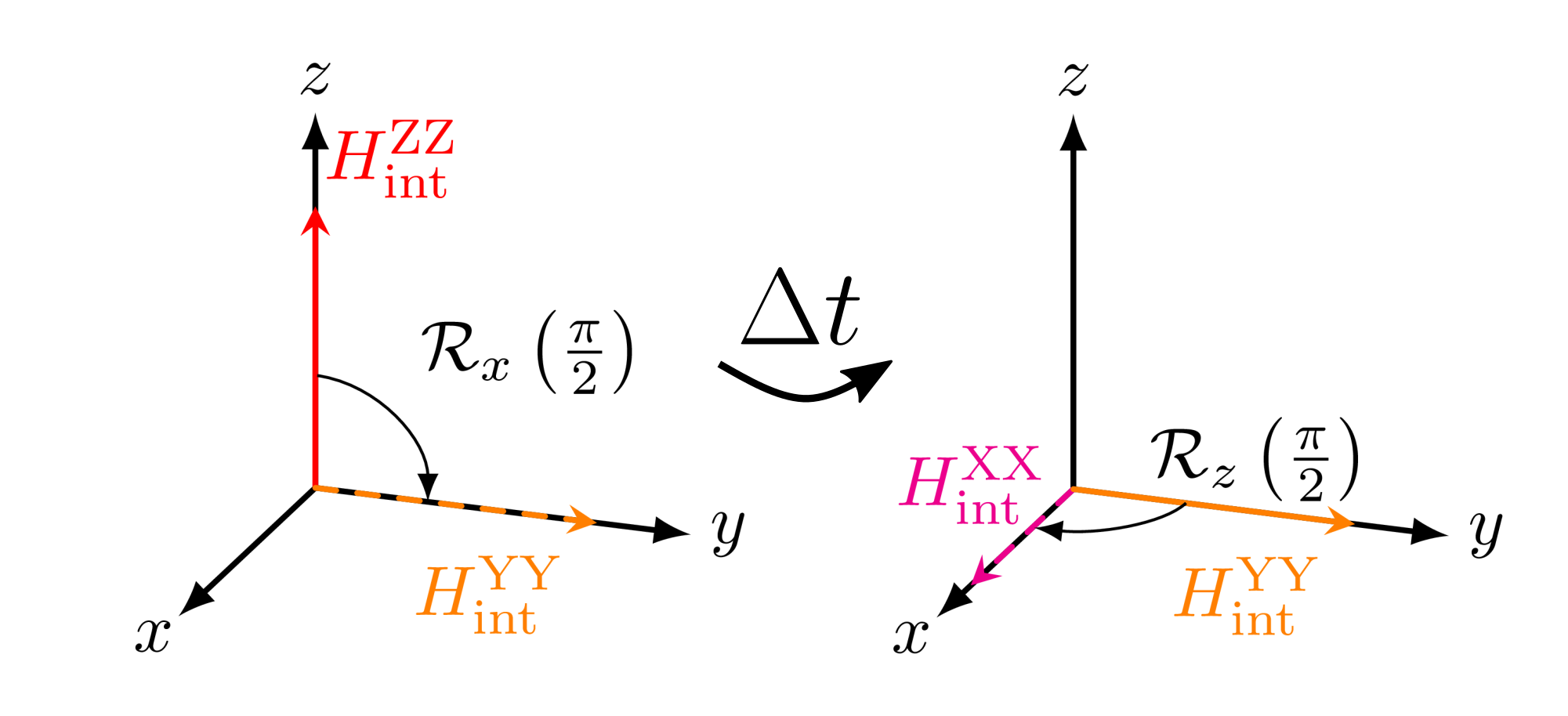}
        \caption{Procedure to obtain an effective dynamical evolution under a Trotterized XYZ Hamiltonian if the considered system only implements a ZZ-type interaction natively using different rotations alternated with free evolutions.}
        \label{fig:digital-analog}
    \end{figure}

Taking the examples that we previously derived as the basis, there are many interesting directions one can pursue:
\begin{itemize}

    \item \emph{Expanding the quantum simulation toolbox by stroboscopic methods} So far, we have considered how to simulate XX and ZZ Hamiltonians. However, as shown e.g., in Refs.~\cite{Heras2014DigitalCircuits,Salathe2015,Huang2016} for qubits, combining such interactions with single qudit rotations one can generate more complex spin-1 Hamiltonians. 
   
    For example, in Fig.~\ref{fig:digital-analog} we show an example of a possible time evolution performed in sequential Trotter steps can be used to engineer the dynamics generated by a full Heisenberg model from an analog ZZ interaction only (up to Trotter error):
    \begin{enumerate}
        \item The system evolves freely under the ZZ Hamiltonian $H^{\text{ZZ}}_{\text{int}}$ for a time $\Delta t$, so the corresponding time-evolution operator for this period is $U_1 = e^{-i H_{\text{int}}^{\text{ZZ}}\Delta t}$.
        \item A single-emitter rotation of $\pi/2$ around the $x$ axis is then applied (using, for example, a pulse) so the $H_{\text{int}}^{\text{ZZ}}$ is mapped to $H_{\text{int}}^{\text{YY}}$ (an interaction along the $y$ axis). Again, the emitters are let to evolve freely under this Hamiltonian for a time $\Delta_t$, so the time-evolution operator is now $U_2 = e^{-i H_{\text{int}}^{\text{YY}}\Delta t}$.
        \item Then, the emitters are rotated $\pi/2$ radians around the $z$ axis, so $H_{\text{int}}^{\text{YY}}\rightarrow H_{\text{int}}^{\text{XX}}$, and evolve for another time $\Delta t$, leading to $U_3 = e^{-i H_{\text{int}}^{\text{YY}}\Delta t}$.
        \item Finally, a rotation $\mathcal{R}_y\left(\frac{\pi}{2}\right)$ is applied to recover the original interaction along the $z$ axis.
    \end{enumerate}
    Thus, the final time-evolution operator is 
    \begin{eqnarray}
        U_\text{XYZ} &&= U_3 U_2 U_1 \approx e^{-i\left(H_{\text{int}}^{\text{XX}}+H_{\text{int}}^{\text{YY}}+H_{\text{int}}^{\text{ZZ}}\right)\Delta t}\nonumber\\
        &&= e^{-i \sum_{i<j} J_{\text{zz}}^{ij} \mathbf{S}_{i}\cdot \mathbf{S}_j\,\Delta t},        
    \end{eqnarray}
    where $J_{\text{zz}}^{ij}$ is given by equation Eq~\ref{eq:Jijzz} in this case. Furthermore, considering a different time-interval in each Trotter step, a general anisotropic spin-1 XYZ Hamiltonian can be simulated.

    \item \emph{Lattice gauge theory quantum simulators}. One of the most attractive avenues nowadays in quantum simulation is to use them to simulate lattice gauge theories (LGT)~\cite{zohar2015,Dalmonte2016,Banuls2020,Zohar2022,Aidelsburger2022,Klco2021}. 
    In particular, this could be applied for the quantum simulation of $\mathbb{Z}_N$-LGTs, which are physically meaningful for two reasons. First, their large $N$-limit reproduces compact QED~\cite{Horn1979}, and thus they may be used as approximations for compact QED~\cite{Kogut1979} which are feasible for quantum simulation. Furthermore, being the center of $SU(N)$ groups, it was shown that $\mathbb{Z}_N$ gauge theories play a key role in confinement effects of such models~\cite{tHooft1978} which is an open and highly important nonperturbative question in particle physics. 
    A challenging aspect in the design of quantum simulators of such theories is that they require tailoring multi-\emph{quNit} quantum gates for the magnetic plaquette interactions. In a recent proposal~\cite{Armon2021}, it was shown how one can harness the photon-mediated interactions to obtain the four-qubit plaquette terms required for $\mathbb{Z}_2$-LGT. The key idea was to use an auxiliary atom to entangle the four neighboring ones, codifying the physical degrees of freedom of the LGT, using the non-local photon-mediated interactions. In Ref.~\cite{Zohar2017b} it was shown using a ZZ interaction between spin-1 atoms is sufficient for implementing the magnetic terms of the $\mathbb{Z}_3$-LGT. Thus, our findings of Sections~\ref{subsec:ZZabstract},\ref{subsec:ZZatoms} can be combined with the proposal of Ref.~\cite{Armon2021} to simulate the most challenging parts of $\mathbb{Z}_3$-LGT Hamiltonians.

    \item \emph{Qudit quantum computation.} Beyond the quantum simulation perspective, our results can also find applications in qudit quantum computation~\cite{Wang2020a}. In fact, the quantum gate analyzed in Fig.~\ref{fig:xx_atomic} is an entangling gate between qudits which can be the basis of more complex ones. Furthermore, if the proper single-qudit gates are given, it is known that a single two-qutrit entangling gate (as the one obtained in this work) is enough to obtain universal qudit quantum computation~\cite{Wang2020a}. Furthermore, from only the XX interaction found here, the exact gate-sequences to obtain universal quantum computation with qutrits are known~\cite{Kempe2002}. Thus, the results in this work can also find applications in qudits quantum computing.
\end{itemize}

\section{Conclusions~\label{sec:conclu}}

Summing up, we have shown how to obtain different type of photon-mediated interaction between spin-1 systems using multi-level emitters. We provide first an emitter-agnostic analysis, and then particularize for the case of atomic emitters where the multi-level structure emerges from hyperfine couplings. In the latter case, we explain how to take care of the complexity introduced by the different Clebsch-Gordan coefficients, and numerically benchmark our results by comparing the effective Hamiltonians obtained with the dynamics obtained by the full system with no approximation. Our results expand the quantum simulation toolbox that can be obtained with cavity and nanophotonic systems to spin-1 models, and can also find applications in qutrit quantum computation. An interesting outlook of our work consist in exploiting larger hyperfine multiplets, like the ones that can be found in Alkaline-Earth atoms~\cite{ludlow15a,daley08a}, to obtain photon-mediated interactions between larger dimensional spin systems.

\begin{acknowledgments}
CT and AGT acknowledge support from CSIC Research   Platform   on   Quantum   Technologies PTI-001, from  Spanish  project PGC2018-094792-B-100(MCIU/AEI/FEDER, EU), and from the Proyecto Sinérgico CAM 2020 Y2020/TCS-6545 (NanoQuCo-CM). EZ acknowledges support by the Israel Science Foundation (grant No. 523/20). 
\end{acknowledgments}
\bibliography{references}
\newpage
\appendix
\begin{widetext}

\section{Multilevel configurations}\label{app:derivations}

In this section, we collect the different multilevel configurations that we propose, showing their full Hamiltonians and calculating the effective Hamiltonians once that the extra degrees of freedom are removed for both the Ising (\ref{app:ZZ}) and the XX (\ref{app:XX}) interactions including the details missed in the main text. In both cases, we divide the discussion into the platform-agnostic one (\ref{app:ZZ_agn} and \ref{app:XX_agn}) and the one where we particularize for atomic systems (\ref{app:ZZ_at} and \ref{app:XX_at}).

\subsection{The spin-1 Ising ZZ interaction}\label{app:ZZ}

\subsubsection{Platform-agnostic case}\label{app:ZZ_agn}

\begin{figure}[h]
    \centering
\includegraphics[width=0.5\linewidth]{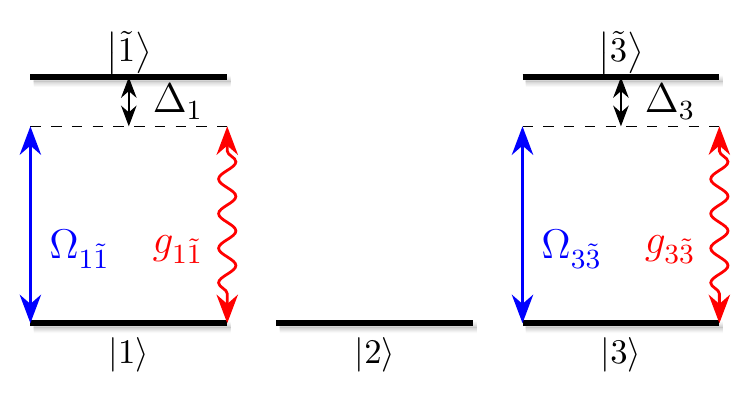}
\caption{Configuration proposed to obtain a spin 1 Ising ZZ Hamiltonian.}
    \label{fig:scheme1}
\end{figure}

This first situation considers two three-level manifolds where a linearly polarized driving connects the ground and the excited states as shown in Fig.~\ref{fig:scheme1}. These excited states can then decay to the original ones through a quantized electromagnetic field, which has the same polarization that the laser.

The Hamiltonian in this case reads (in a frame rotating with the laser frequency $\omega_L$):
\begin{equation}\label{eq:initial_ZZ_agnostic}
    H = H_\text{s}+H_\text{f} +H_L + H_I
\end{equation}
Here $H_\text{s}$ is the projection of the Hamiltonian over the slow/ground state subspace,
\begin{equation}\label{eq:zz_agn_fast}
    H_\text{s}=\sum_{i}\Delta_1\s{\tilde{1}}{\tilde{1}}{i}+\Delta_2 \s{\tilde{2}}{\tilde{2}}{i} + \Delta_{3}\s{\tilde{3}}{\tilde{3}}{i}+\Delta_a a^\dagger a
\end{equation}
accounts for the non-interacting parts of both the excited emitters states and the field, with a single driving frequency $\omega_L$ and detunings $\Delta_i= \omega_i -\omega_L $ and $\Delta_a=  \omega_a-\omega_L$, with $\omega_a$ being the cavity mode energy. On the other hand, the laser field enters $H$ as an external driving
\begin{equation}
    H_L = \sum_{i}\frac{\om{1}}{2}\s{\tilde{1}}{1}{i}+\frac{\om{3}}{2}\s{\tilde{3}}{3}{i}+\mathrm{H.c.}
\end{equation}
meanwhile the light-matter Hamiltonian reads:
\begin{equation}
    H_I = \sum_{i}\g{1}^i \s{1}{\tilde{1}}{i}a^\dagger+\g{3}^i\s{3}{\tilde{3}}{i}a^\dagger+\mathrm{H.c.}
\end{equation}
(note that we do not include a term for the $\ket{2}\leftrightarrow\ket{\tilde{2}}$ transition). Finally, the Linblad operators account for spontaneous emission from each atomic excited state, $L_{\gamma,1} = \sum_i \sqrt{\gamma_1}\s{1}{1'}{i}$ and $L_{\gamma,3} = \sum_i \sqrt{\gamma_3}\s{3}{3'}{i}$; but also for the lose of photons, $L_\kappa = \sqrt{\kappa}a$.

Following now the procedure described in Sec~.\ref{sec:system} of the main text, we can eliminate first the excited emitter states, leading to the following renormalized light-matter Hamiltonian:
\begin{eqnarray}
    H_\text{eff}^{(1)} &&= H_s + \Delta_a a^\dagger a -\sum_i\Bigg[\frac{\Delta_1|\om{1}|^2}{4\Delta_1^2 + \gamma_1^2}\s{1}{1}{i}-\frac{\Delta_3|\om{3}|^2}{4\Delta_3^2 + \gamma_3^2}\s{3}{3}{i}-\left(\frac{4\Delta_1 |g_{1\tilde{1}}^{i}|^2}{4\Delta_1^2+\gamma_1^2}\s{1}{1}{i}+\frac{4\Delta_1 |g_{3\tilde{3}}^{i}|^2}{4\Delta_3^2+\gamma_3}\s{3}{3}{i}\right)a^\dagger a\nonumber\\
    &&-\left(\frac{2\om{1}\gc{1}^i\Delta_1}{4\Delta_1^2+\gamma_1^2}\s{1}{1}{i}+\frac{2\om{3}\gc{3}^i\Delta_3}{4\Delta_3^2+\gamma_3^2}\s{3}{3}{i}\right)a^\dagger\Bigg]+ \mathrm{H.c.}
\end{eqnarray}
We want to find an effective description that only involves the atomic operators. Thus, we further assume that a large cavity detuning so that its population is very small, and can be adiabatically eliminated. Doing that, we arrive to the following effective Hamiltonian:
\begin{eqnarray}\label{eq:effZZ_appendix}
    H_\mathrm{eff} &&= H_s-\sum_{i}\left[\frac{\Delta_1|\om{1}|^2}{4\Delta_1^2 + \gamma_1^2}\s{1}{1}{i}+\frac{\Delta_3|\om{3}|^2}{4\Delta_3^2 + \gamma_3^2}\s{3}{3}{i}+
    \frac{4\Delta_a}{4\Delta_a^2+\kappa^2}\left(\mu_i|^2\s{1}{1}{i}+|\nu_i|^2\s{3}{3}{i}\right)\right]+H_\text{int}\,,
\end{eqnarray}
where we neglect the terms proportional to $a^\dagger a$ (since the population of the field mode is assumed to be small under the conditions of validity of this elimination), and introduce the parameters:
\begin{equation}\label{eq:scheme1_munu}
        \mu_i = \frac{2\om{1}\gc{1}^i\Delta_1}{4\Delta_1^2+\gamma_1^2}\,\,\text{and}\,\,\nu_i = \frac{2\om{3}\gc{3}^i\Delta_3}{4\Delta_3^2+\gamma_3^2}\,.
\end{equation}

Here, the term $H_\text{int}$ is the one responsible for the photon-mediated interactions between emitters which reads:
\begin{eqnarray}
    H_\text{int}=-\frac{4\Delta_a}{4\Delta_a^2+\kappa^2}\sum_{i\neq j}&&\left(\mu_i\bar{\mu}_j\s{1}{1}{i}\s{1}{1}{j}+\nu_i\bar{\nu}_j\s{3}{3}{i}\s{3}{3}{j}+\mu_i\bar{\nu_j}\s{1}{1}{i}\s{3}{3}{j}+\bar{\mu_i}\nu_j\s{1}{1}{j}\s{3}{3}{i}\right)\,.
\end{eqnarray}

Let us now consider that we can tune the Hamiltonian parameters such that $\mu_i=-\nu_i$ for each atom $i$. This situation can be achieved by several means: for example, setting $\om{1}=\om{-1}$, $\g{1}^i=-\g{-1}^i$, $\gamma_1=\gamma_3$ and $\Delta_1=\Delta_3$, or instead changing the relative sign between the coupling constants or the driving amplitudes. Anyway, once the condition $\mu_i = -\nu_i$ is satisfied, the effective Hamiltonian after the first adiabatic elimination derived above, $H_\text{eff}^{(1)}$, reads:
\begin{equation}
    H_\text{eff}^{(1)} = -\sum_i \mu_i \left(\s{1}{1}{i}-\s{3}{3}{i}\right)a^\dagger + \text{H.c.}+\ldots = -\sum_i \mu_i S_{z}^{i} a^\dagger + \text{H.c.}+\ldots\,,
\end{equation}
where we have only written the $S_z^i \equiv \s{1}{1}{i}-\s{3}{3}{i}$ terms generating the ZZ interactions. Hence, once the quantized field is eliminated, these terms proportional to $S_{z}^{i}$ will generate the desired $S_{z}^{i}S_{z}^{j}$ interactions:
\begin{equation}\label{eq:scheme1_ising}
   H_\text{eff} = H_s + \sum_{i<j} J_\mathrm{z}^{ij} S^{i}_z S^{j}_z\,,
\end{equation}
with $J_{\mathrm{z}}^{ij} = \Re\left[\frac{-8\Delta_a\mu_i \bar{\mu}_j }{4\Delta_a^2+\kappa^2}\right]$. Furhermore, the condition $\mu_i = -\nu_i$ obtained as discussed above yields a pair of AC Stark shifts for the $\ket{1}$ and $\ket{3}$ levels in the effective Hamiltonian of Eq.~\ref{eq:effZZ_appendix} that are equal in amplitude and sign, so they can be easily removed with a single laser field coupled off-resonantly to a different transition. Hence, the effective Hamiltonian in Eq.~\ref{eq:scheme1_ising} corresponds to an Ising-type interaction for the emitters. As a final remark, let us note that although the $S_{z}^{i}$ operator only has two matrix elements different from zero (and these have an opposite sign), it \emph{cannot} be mapped to a spin 1/2 projection along the $z-$axis, $\sigma_{z}^{i}$. This is because the identity in this spin 1/2 space would be $\mathbb{1}_{s=1/2}=\sigma_{11}+\sigma_{22}=\sigma_{z}^2$, meanwhile the identity in the spin 1 space is $\mathbb{1}_{s=1}=\sigma_{11}+\sigma_{00}+\sigma_{11}\neq S_{z}^2$.

\subsubsection{Implementing the interaction with real atoms}\label{app:ZZ_at}

The configuration proposed in the main text to obtain an effective ZZ interaction between two Alkali atoms uses the $D_1$ line of Rubidium (or Sodium) and an extra microwave field to induce two-photon transitions. The Hamiltonian, now in a frame rotating with the frequency of the laser and also of the microwave field, reads:
\begin{equation}
    H = H_\text{s}+H_\text{f} +H_L + H_I + H_m
\end{equation}
Again $H_\text{s}$ is the projection of the Hamiltonian over the slow subspace,
\begin{equation}\label{eq:zz_at_fast}
    H_\text{f}=\sum_{i}\Delta_1\left(\s{\tilde{1}}{\tilde{1}}{i,\tilde{F}=1}+\s{\tilde{0}}{\tilde{0}}{i,\tilde{F}=1}+\s{-\tilde{1}}{-\tilde{1}}{i,\tilde{F}=1}\right)+\Delta_2 \left(\s{\tilde{1}}{\tilde{1}}{i,\tilde{F}=2}+\s{\tilde{0}}{\tilde{0}}{i,\tilde{F}=2}+\s{-\tilde{1}}{-\tilde{1}}{i,\tilde{F}=2}\right)+\Delta_a a^\dagger a
\end{equation}
accounts for the non-interacting parts of both the excited hyperfine angular momentum $\tilde{F}=1$ and $\tilde{F}=2$ and also the field, with an optical driving frequency $\omega_L$ and a microwave one $\omega_{MW}$, that yield the detunings $\Delta_1= \omega_1 -\omega_L +\omega_{MW}$, $\Delta_2 = \omega_2 -\omega_L$ and $\Delta_a=  \omega_a-\omega_L + \omega_{MW}$. Furthermore, the laser field induces a driving term that is coupled only to the $\tilde{F}=2$ line,
\begin{equation}
    H_L = \sum_{i}\frac{\om{1}}{2}\s{\tilde{1}}{1}{i,\tilde{F}=2}+\frac{\om{0}}{2}\s{\tilde{0}}{0}{i,\tilde{F}=2}+\frac{\Omega_{-1\tilde{1}}}{2}\s{-\tilde{1}}{-1}{i,\tilde{F}=2}+\mathrm{H.c.}\,,
\end{equation}
the light-matter Hamiltonian is
\begin{equation}
    H_I = \sum_{i}\g{1}^i \s{1}{\tilde{1}}{i,\tilde{F}=1}a^\dagger+g_{-1\tilde{1}}^i\s{-1}{-\tilde{1}}{i,\tilde{F}=1}a^\dagger+\mathrm{H.c.}
\end{equation}
and the microwave driving inducing transitions between the $\tilde{F}=2$ and the $\tilde{F}=1$ states reads:
\begin{equation}
    H_m = \sum_i \frac{\Omega^\text{MW}_{1}}{2}|\tilde{F}=1,1\rangle_i\langle\tilde{F}=2,1|+\frac{\Omega^\text{MW}_{0}}{2}|\tilde{F}=1,0\rangle_i\langle\tilde{F}=2,0|+\frac{\Omega^\text{MW}_{-1}}{2}|\tilde{F}=1,-1\rangle_i\langle\tilde{F}=2,-1|+\text{H.c.}
\end{equation}
(where we have recovered the full braket notation to highlight that the transitions always take place among the excited atomic states). Hence, the first step is the elimination of the $\tilde{F}=2$ levels, leading to the following effective Hamiltonian (in the following, we omit the $\tilde{F}=1$ superscript so $\s{k}{l}{i,\tilde{F}=1}\equiv \s{k}{l}{i}$):
\begin{eqnarray}\label{eq:effZZ_atomic_noMW}
    H_\text{eff}^{\tilde{F}=1} = H_\text{s}+ \sum_{i,j} \left(\Delta_1 -\frac{\Delta_2 \left|\Omega_j^\text{MW}\right|^2}{4\Delta_2^2+\gamma_2^2}\right)\s{\tilde{j}}{\tilde{j}}{i}+\left(\frac{\Omega_j^\text{eff}}{2}\s{j}{\tilde{j}}{i}+\g{1}^i \s{1}{\tilde{1}}{i}a^\dagger+g^i_{-1\tilde{1}}\s{-1}{-\tilde{1}}{i}a^\dagger+\mathrm{H.c.}\right)+\Delta_a a^\dagger a \,,
\end{eqnarray}
with $\Omega^{\text{eff}}_{j}\equiv -2\Omega_{j\bar{j}}\Omega^\text{MW}_{j}\Delta_2/(4\Delta_{2}^{2}+\gamma_{2}^2)$ and where $\gamma_2$ is the rate of spontaneous emission from that level. We find that the elimination of the $\tilde{F=2}$ excited levels has two effects: on the one hand, it introduces a state-dependent AC Stark shift on the $\tilde{F}=1$ line, depending on the amplitude of the microwave driving and the detuning $\Delta_2$; and, on the other hand, it renormalizes the driving amplitudes to consider the two-photon transitions with effective amplitude given by $\Omega_j^\text{eff}$. The key aspect of this scheme is that the effective two-photon Raman transitions from the $F=1$ level to the $\tilde{F}=1$ once that the $\tilde{F}=2$ level has been eliminated have the same sign for both the $m_F=1$ and $m_F = -1$ cases, so the remaining adiabatic elimination is able to exploit the fact that $g^i_{1\tilde{1}}$ and $g^i_{-1\tilde{1}}$ have opposite signs. 

Thus, once that the extra $\tilde{F}=2$ levels are eliminated, the Hamiltonian in Eq.~\ref{eq:effZZ_atomic_noMW} is identical to the one in Eq.~\ref{eq:initial_ZZ_agnostic}, so the procedure to derive the final effective Hamiltonian would be equal: the only difference apart from the state dependent AC Stark shifts entering the detunings and the renormalized driving amplitudes discussed above would be an extra term in the final AC stark shift,
\begin{equation}\label{eq:ZZ_at_Stark}
    H_\text{Stark} = \sum_{i,j}\frac{|\Omega_\text{eff}^{j}|^2 \Delta_{1,j}^{\text{eff}}}{4\Delta_{1,j}^{\text{eff }2}+\gamma_{1}^{2}}\s{j}{j}{i}=\sum_{i,j}\delta_j \s{j}{j}{i}\,,
\end{equation}
where we have introduced the effective detunings 
\begin{equation}
    \Delta^{\text{eff}}_{1,j} = \Delta_1 -\frac{\Delta_2 \left|\Omega_j^\text{MW}\right|^2}{4\Delta_2^2+\gamma_2^2}\,.
\end{equation}
Finally, using the fact that $|\Omega_\text{eff}^1| = |\Omega_\text{eff}^{-1}|$ and $\Delta_{1,1}^{\text{eff}}=\Delta_{1,-1}^{\text{eff}}$, so $\delta_1 = \delta_{-1}$, we can rewrite the Stark shift in Eq.~\ref{eq:ZZ_at_Stark} as $H_\text{Stark} = \sum_{i} \delta_0 \mathbb{1}_i++\left(\delta_1-\delta_0\right)S_{z}^{i\,2}$, neglect the term proportional to the unit operator $\mathbb{1}_i$ and finally arrive to the Hamiltonian written in the main text,
\begin{eqnarray}
    H_\text{Stark} =\delta \sum_i  S_{z}^{i\,2}\,,
\end{eqnarray}
where $\delta \equiv\delta_1 - \delta_0$ is the amplitude of the shift.

\subsection{The spin 1 XX exchange interaction}\label{app:XX}

\subsubsection{Platform-agnostic case}\label{app:XX_agn}

\begin{figure}[h]
    \centering
\includegraphics[width=0.5\linewidth]{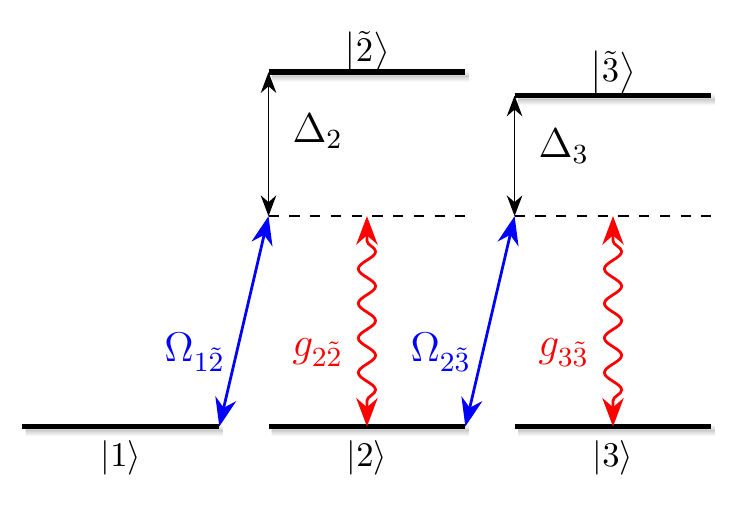}
\caption{Configuration used to obtain an effective XX-type interaction between the emitters using a single field mode.}
    \label{fig:scheme3}
\end{figure}

In this Appendix we consider crossed polarizations for the first time. This will couple different states in the atomic operators, leading to terms of the form $\sigma_{\alpha \beta}^i$ before the field is eliminated (with $\alpha\neq\beta$) that, in turn, will lead to spin-exchange like interactions. A model with the minimal ingredients to obtain this is shown in Fig.~\ref{fig:scheme3}, with a circularly polarized laser driving generating the transitions $\ket{1}\leftrightarrow\ket{2'}$ and $\ket{2}\leftrightarrow\ket{3'}$ (the configuration with the $\ket{3}\leftrightarrow\ket{2'}$ and $\ket{2'}\leftrightarrow\ket{1}$ transitions would be equivalent, so we only take into account one in the discussion). The Hamiltonian describing this situation is again $H= H_\mathrm{s}+H_\mathrm{f}+H_L+H_I$, and we find that $H_\mathrm{f}$ is given again by Eq.~\ref{eq:zz_agn_fast} meanwhile the driving Hamiltonian now is
\begin{equation}
    H_L = \sum_{i}\frac{\oms{2}{1}}{2}\s{1}{\tilde{2}}{i}+\frac{\oms{3}{2}}{2}\s{2}{\tilde{3}}{i}+\mathrm{H.c.}
\end{equation}
and the light-matter interaction term reads:
\begin{equation}
    H_I=\sum_{i} \g{1}^i\s{1}{\tilde{1}}{i}a^\dagger+
    \g{2}^i\s{2}{\tilde{2}}{i}a^\dagger+\g{3}^i\s{3}{\tilde{3}}{i}a^\dagger+\mathrm{H.c.}\,,
\end{equation}
both already written in a rotating frame with the single laser frequency $\omega_L$ (note that here we have written explicitly the $1\leftrightarrow\tilde{1}$ although as we will show below, it will not play a role in the dynamics with this laser configuration and the assumption of a far detuned cavity).

Finally, the Linblad operators account for spontaneous emission from each atomic excited state (via the two channels), $L_{\gamma_1,1} = \sum_i \sqrt{\gamma_1}\s{1}{\tilde{1}}{i}$, $L_{\gamma_2,1}=\sum_i\sqrt{\gamma_2}\s{1}{\tilde{2}}{i}$, $L_{\gamma_2,2}=\sum_i \sqrt{\gamma_2}\s{2}{\tilde{2}}{i}$, $L_{\gamma_3,1} = \sum_i \sqrt{\gamma_3}\s{2}{\tilde{3}}{i}$ and $L_{\gamma_3,2}=\sum_i \sqrt{\gamma_3}\s{3}{\tilde{3}}{i}$, but also for the lose of photons $L_\kappa = \sqrt{\kappa}a$.

Now, following the procedure described in Sec.~\ref{sec:system}, we can eliminate first the excited emitter states and find the following Hamiltonian:
\begin{eqnarray}\label{eq:xx_appendix_pre}
    H_\text{eff}^{(1)} = H_\text{s} -\sum_{i}&&\Bigg\{\frac{\Delta_2|\oms{2}{1}|^2}{4\Delta^2_2+4\gamma_2^2}\s{1}{1}{i}+\frac{\Delta_3|\oms{3}{2}|^2}{4\Delta^2_3+4\gamma_3^2}\s{2}{2}{i}
    +\left(\frac{4\Delta_1 |\gs{1}{1}^i|^2}{4\Delta_1^2+\gamma_1^2}\s{1}{1}{i}+\frac{4\Delta_2 |\gs{2}{2}^i|^2}{4\Delta_2^2+4\gamma_2^2}\s{2}{2}{i}+\frac{4\Delta_2 |\gs{2}{3}^i|^2}{4\Delta_2^2+4\gamma_2^2}\s{3}{3}{i}\right)a^\dagger a\nonumber\\
    +&&\Big[\left(\frac{2\oms{2}{1}\gs{2}{2}^i\Delta_2}{4\Delta_2^2+4\gamma_2^2}\s{2}{1}{i}+\frac{2\oms{3}{2}\gs{3}{3}^i\Delta_3}{4\Delta_3^2+4\gamma_3^2}\s{3}{2}{i}\right)a^\dagger + \mathrm{H.c.}\Big]\Bigg\}
\end{eqnarray}
(note that this Hamiltoinian already takes into account the two possible channels of decay to the $\ket{2}$ and $\ket{3}$ states, meanwhile the $\ket{1}$ state only has one: hence the difference in the corresponding terms). Furthermore, since we are looking for a Hamiltonian describing photon-mediated interactions between the emitters, we further suppose a situation where the cavity field is far detuned and therefore it has a very small population, and we can neglect the terms proportional to $a^\dagger a$ (so all the contributions proportional to $g_{1\tilde{1}}$ vanish, and this is the reason why this transition is already neglected in the main text). Furthermore, introducing the parameters
\begin{equation}
    \xi_i = \frac{2\oms{2}{1}\gs{2}{2}^i\Delta_2}{4\Delta_2^2+4\gamma_2^2}\quad\text{and}\quad \eta_i = \frac{2\oms{3}{2}\gs{3}{3}^i\Delta_3}{4\Delta_3^2+4\gamma_3^2}
\end{equation}
and eliminating the cavity field finally yields the effective Hamiltonian:
\begin{eqnarray}\label{eq:app_xx_eff_agn}
    H_\mathrm{eff} &&= H_\text{s} -\sum_{i}\Big(\frac{\Delta_2|\oms{2}{1}|^2}{4\Delta^2_2+4\gamma_2^2}\s{1}{1}{i}+\frac{\Delta_3|\oms{3}{2}|^2}{4\Delta^2_3+4\gamma_3^2}\s{2}{2}{i}+\frac{4\Delta_a\left|\xi_i\right|^2}{4 \Delta_a^2+\kappa^2}\s{1}{1}{i}+\frac{4\Delta_a\left|\eta_i\right|^2}{4 \Delta_a^2+\kappa^2}\s{2}{2}{i}\Big)+H_\text{int}
\end{eqnarray}
found in the main text, where the term $H_\text{int}$ reads:
\begin{eqnarray}\label{eq:int_XX_int_app}
    H_\text{int} = -\frac{4\Delta_a}{4\Delta_a^2+\kappa^2}&&\sum_{i\neq j}\Big[\xi_i\bar{\xi_j}\s{2}{1}{i}\s{1}{2}{j}+\eta_i\bar{\eta_j} \s{3}{2}{i}\s{2}{3}{j}+\xi_i\bar{\eta_j}\s{2}{1}{i}\s{2}{3}{j}+\bar{\xi_j}\eta_i\s{3}{2}{i}\s{3}{2}{j}\Big]\,.
\end{eqnarray}
To understand clearly the reason why the condition $\xi_i = \eta_i$ yields an XX interaction, note that we can rewrite the excitation operator used to find this final Hamiltonian from the one in Eq.~\ref{eq:xx_appendix_pre} (according to the procedure in Sec.~\ref{sec:system}) as
\begin{equation}
    V_+ = -\sum_i\left(\xi_i\s{2}{1}{i}+\eta_i\s{3}{2}{i}\right)a^\dagger\,.
\end{equation}
If we could be able to tune the system parameters in such a way that $\xi_i=\eta_i$, then the operator above would be
\begin{equation}
    V_+ = -\sum_i\frac{\xi_i}{\sqrt{2}}\left(\sqrt{2}\s{2}{1}{i}+\sqrt{2}\s{3}{2}{i}\right)a^\dagger =-\sum_i\frac{\xi_i}{\sqrt{2}} S_+^i a^\dagger\,,
\end{equation}
where we have introduced the ladder operator for the $i-$th atom,
\begin{equation}
    S_+^i = \sqrt{2}\left(\s{2}{1}{i}+\s{3}{2}{i}\right)\,,
\end{equation}
with $S_-^i = \left(S_+^i\right)^\dagger$. Then, for different emitters (with $i\neq j$) the adiabatic elimination of the field would yield terms of the form
\begin{equation}\label{eq:pm}
    H_\text{int}=\frac{2|\xi_i|^2\Delta_a}{4\Delta_{a}^2+\kappa^2}\sum_{i\neq j} S_{+}^i S_{-}^j = \frac{2|\xi_i|^2\Delta_a}{4\Delta_{a}^2+\kappa^2}\sum_{i<j}\left(S_{+}^i S_{-}^j + S_{-}^i S_{+}^j\right)\,.
\end{equation}
Since the ladder operators can be written in terms of the $S^i_x$ and $S^i_y$ operators as
\begin{equation}
    S^i_\pm = S_x^i \pm i S_y^i\,,
\end{equation}
the term in equation \eqref{eq:pm} can be casted as
\begin{eqnarray}\label{eq:scheme3xx}
    H_\text{int}=\sum_{i<j}J_{\text{xx}}^{ij}\left(S_x^i S_x^j + S_y^i S_y^j\right)\,,
\end{eqnarray}
with 
\begin{eqnarray}
    J_\text{xx}^{ij}=\Re\left[\frac{-4\xi_i\bar{\eta_i}\Delta_a}{4\Delta_a^2+\kappa^2}\right]\,,
\end{eqnarray}
The interaction in equation \eqref{eq:scheme3xx} is an XX spin model with a coupling strength given by $J_{\text{xx}}^{ij}$.

\subsubsection{Implementing the interaction with real atoms}\label{app:XX_at}

In the main text we proposed the hyperfine levels of Alkali atoms, in particular, the D2 line of Rubidium or Sodium, as a suitable physical platform to implement the proposed models. The full Hamiltonian describing this situation would be 
\begin{equation}
    H = H_\text{s}+H_\text{f} +H_L + H_I + H_m\,,
\end{equation}
where $H_\text{s}$ is the projection of the Hamiltonian over the slow subspace, $H_\text{f}$ is the Hamiltonian in Eq.~\ref{eq:zz_at_fast} accounting for the excited $\tilde{F}=1$ and $\tilde{F}=2$ states and also the field (and where the same global rotations have been applied to make it time-independent, producing the detunings $\Delta_1 \equiv \omega_1 - \omega_L + \omega_{MW}$, $\Delta_2\equiv \omega_2 - \omega_L$ and $\Delta_a = \omega_a - \omega_L + \omega_{MW}$, were $\omega_L$ is the laser frequency and $\omega_{MW}$ is the frequency of the microwave field) the laser field enters as a classical driving term $H_L$ coupling the $\tilde{F}=2$ states and the $F=1$ ones,
\begin{equation}
    H_L = \sum_{i}\left[\frac{\Omega_{1\tilde{0}}}{2}\s{\tilde{0}}{1}{i,\tilde{F}=2}+\frac{\Omega_{-0\tilde{1}}}{2}\s{\tilde{0}}{-1}{i,\tilde{F}=2}+\frac{\Omega_{-1\tilde{2}}}{2}\s{-\tilde{2}}{-1}{i,\tilde{F}=2}\right]+\mathrm{H.c.}\,,
\end{equation}
the light-matter Hamiltonian reads
\begin{equation}
    H_I = \sum_{i}\left[\g{0}^i \s{0}{\tilde{0}}{i,\tilde{F}=2}a^\dagger+g_{-1\tilde{1}}^i\s{-1}{-\tilde{1}}{i,\tilde{F}=2}a^\dagger\right]+\mathrm{H.c.}
\end{equation}
and the microwave driving inducing transitions between the $\tilde{F}=2$ and the $\tilde{F}=1$ states reads:
\begin{equation}
    H_m = \sum_i \frac{\Omega^\text{MW}_{1}}{2}|\tilde{F}=1,1\rangle_i\langle\tilde{F}=2,1|+\frac{\Omega^\text{MW}_{0}}{2}|\tilde{F}=1,0\rangle_i\langle\tilde{F}=2,0|+\frac{\Omega^\text{MW}_{-1}}{2}|\tilde{F}=1,-1\rangle_i\langle\tilde{F}=2,-1|+\text{H.c.}
\end{equation}
To relate this situation with the one derived in the previous section, we can first eliminate the excited $\tilde{F}=1$ levels. Note that we are not coupling to the $\tilde{F}=1$ level either the driving term nor the field one, so the only effect of the elimination of these levels is a renormalization of the detunings in the excited state Hamiltonian. Hence, the first effective Hamiltonian without the $\tilde{F}=1$ states reads (we will omit the $\tilde{F}=2$ superscript in the following and assume that all the coherence operators refer to this manifold, $\s{k}{l}{i,\tilde{F}=2}\equiv \s{k}{l}{i}$):
\begin{eqnarray}\label{eq:effXX_atomic_noMW}
    H_\text{eff}^{\tilde{F}=2} = H_\text{s}+ \sum_{i,j}\Delta^{\text{eff}}_{2,j}\s{\tilde{j}}{\tilde{j}}{i}+\Delta_a a^\dagger a +H_L + H_I \,,
\end{eqnarray}
where we have introduced the state-dependent energy shifts $\Delta^{\text{eff}}_{2,j} =  \Delta_2 -\frac{\left|\Omega_j^\text{MW}\right|^2}{4\Delta_1^2}$. The only difference between the Hamiltonian in Eq.~\ref{eq:effXX_atomic_noMW} and the full Hamiltonian before the eliminations used in the Appendix~\ref{app:XX_agn} to derive the XX model is a state-dependent energy shift in $H_\text{f}$ according to $\Delta^{\text{eff}}_{2,j}$. This implies that the parameters $\xi_i$ and $\eta_i$ now would be:
\begin{equation}
    \xi_i = \frac{\oms{2}{1}\gs{2}{2}^i}{2\Delta^{\text{eff}}_{2,0}}\quad\text{and}\quad \eta_i = \frac{2\oms{3}{2}\gs{3}{3}^i}{2\Delta^{\text{eff}}_{2,-1}}\,,
\end{equation}
and the condition $\xi_i = \eta_i$, substituting the values of the Clebsch-Gordan coefficients in tables \ref{tab:CG_sigma-} and \ref{tab:MW_pi} in the main text, yields the an equivalent condition over the renormalized detunings, $3\Delta^{\text{eff}}_{2,0}=2\Delta^{\text{eff}}_{2,-1}$, that can be solved for
\begin{equation}
       \left|\Omega^\text{MW}\right|=\sqrt{\frac{8}{3}\Delta_1 \Delta_2}\,.
\end{equation}
Finally, let us remark that together with the interaction Hamiltonian derived in the Appendix \ref{app:XX_agn}, Eq.~\ref{eq:scheme3xx}, the full effective Hamiltonian in Eq.~\ref{eq:app_xx_eff_agn} also includes terms that in this atomic case would be proportional to the $\sigma_{11}^{i}$ and $\sigma_{00}^{i}$ operators, together with an extra one proportional to $\sigma_{-1-1}^{i}$ that could be grouped in a Hamiltonian:
\begin{equation}
    H_\text{Stark}= \sum_{i,j}\epsilon_j \s{j}{j}{i}\,,
\end{equation}
with $\epsilon_i = \left|C_i\right|^2 \Omega^2/\Delta^{\text{eff}}_{2,i}$, that can be casted in terms of the unit matrix and the operators $S_z^{i}$ and $S_{z}^{i\,2}$ as
\begin{equation}
    H_\text{Stark} = \sum_i \epsilon_0 \mathbb{1}^i+\frac{\epsilon_1 - \epsilon_{-1}}{2}S_{z}^{i} + \frac{\epsilon_1 + \epsilon_{-1}-2\epsilon_0}{2}S_{z}^{i\,2}\,.
\end{equation}

\section{Scaling of the errors}\label{app:errors}

In the main text (Sec.~\ref{subsubsec:ZZlosses}) we claimed that there are two possible sources of error when comparing the full quantum evolution with the corresponding effective one:
\begin{itemize}
    \item First, there is a trade-off between the coherent proccesses (according to the timescale $\sim J^{ij}_{\text{z}}$ for the ZZ interaction and $\sim J^{ij}_{\text{xx}}$ for the XX interaction) and the corresponding losses, leading to an error $\epsilon_{\text{losses}}$ that scales as 
    \begin{equation}\label{eq:epsilon_losses_app}
        \epsilon_{\text{losses}}\sim \frac{1}{\sqrt{C}}    
    \end{equation}
    if the detuning $\Delta_a$ is optimized according to Eq.~\eqref{eq:optimal_det}. This is an error due to the losses, that is present even in the non-projected case (when comparing the actual dynamics with the purely coherent ones), hence the name.
    
    \item And second, there is an error (even in an ideal coherent case, without any loss) found when comparing the effective quantum evolutions with the full ones, since the effective Hamiltonians are obtained perturbatively according to Eq.~\eqref{eq:theory_effH} and thus the formula is not valid if the perturbative correction is big enough, that is, if the slow subspace cannot be regarded as slow when compared with the other timescales of the system. Thus, when the condition above is used to make $\epsilon_{\text{losses}}\rightarrow0$ as $C\rightarrow \infty$, the coupling constant between the emitters and the field $g$ cannot be arbitrarily high, or otherwise this perturbative calculation would fail. The order of this correction is $J\propto \Omega^2 g/\Delta^2 \Delta_a$, and it is necessary that this energy scale is much smaller than the ones typical in the excited spaces, $\Delta\approx \Delta_a$. Hence, we quantify the error of the approximation in terms of 
    \begin{equation}\label{eq:epsilon_approx_app}
        \epsilon_\text{approx} \sim J/\Delta\sim \frac{\left|\Omega\right|^2}{\Delta^2} \frac{|g|^2}{\Delta_{a}\Delta}\,.
    \end{equation}
\end{itemize}

Hence, the total error between the effective photon-mediated Hermitian evolution according to the considered spin models (such as the ZZ or the XX) and the actual evolution (including both the non-Hermitian jump operators and the excited levels) can be quantified in terms of the infidelity $\mathcal{I}$ between states evolved in these two ways, see Eq.~\eqref{eq:infidelity_def} of the main text and the accompanying discussion. Note that this infidelity will include all the possible sources of error, beyond the ones considered here (that are obtained from series expansion up to first order in $\gamma/\Delta$ and $\kappa/\Delta_a$), so $\mathcal{I}=\epsilon_\text{losses}+\epsilon_\text{approx}+\dots$.

Furthermore, we found that the condition of optimal detuning to minimize the losses in Eq.~\eqref{eq:optimal_det} sets $\epsilon_{\text{approx}}=|\Omega|^2\sqrt{C}\gamma/\Delta^3$. Recalling that the ratio $\Omega/\Delta$ is independent of the values of the cooperativity, but should not be extremely small if a fast enough evolution is desired, we assumed it to be fixed (moreover, the value of $\Omega/\Delta$ does not affect the single-atom cooperativity $C$). Hence, $\epsilon_{\text{approx}}$ goes to infinity in the limit $C\rightarrow\infty$ if the ratio $\gamma/\Delta$ is not kept low (or conversely, $\Delta/\gamma$ big). Therefore, in Sec.~\ref{subsubsec:ZZlosses} we studied these two situations and we found numerical results that corroborate these claims, as shown in Fig.~\ref{fig:zz_agnostic}. In this Appendix we further elaborate this numerical results.

\begin{figure}[t]
    \centering
    \includegraphics[width=0.9\linewidth]{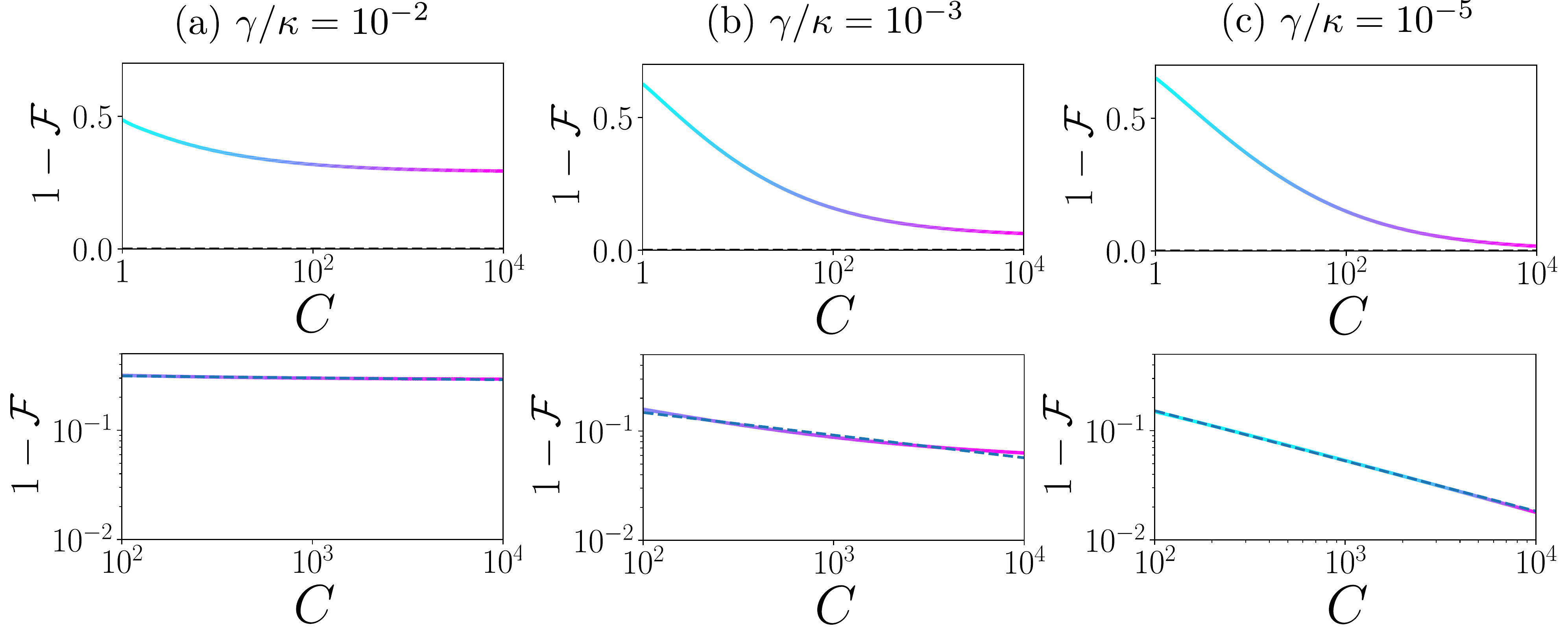}
    \caption{(a-c) Infidelity (Eq.~\eqref{eq:infidelity_def}) as function of the cooperativity for different values of $\gamma/\kappa$, that introduces a bigger error in the approximation as this ratio is increased according to Eq.~\eqref{eq:eps_approx_app}. The figures reproduce the results shown in the top panel of Fig.~\ref{fig:zz_agnostic}(c) of the main text, so the parameters taken are equal to the ones there with the except of the ratio $\gamma/\kappa$. Furthermore, the linear fits superimposed over the loglog plots in the bottom panels yield slopes of $-0.02$ (a), $-0.21$ (b) and $-0.46$ (c).}
    \label{fig:appB_coop}
\end{figure}

Thus, in Fig.~\ref{fig:appB_coop} we have reproduced the top panel of Fig.~\ref{fig:zz_agnostic}(c) in the main text for different values of the ratio $\gamma/\kappa$ but keeping the same values of the cooperativity. This makes $g/\Delta_a$ bigger as the ratio increases, so although Fig.~\ref{fig:zz_agnostic}(c) is used in the main text mainly to study the effect $\epsilon_{\text{losses}}$, the error in the approximation can be casted as
\begin{equation}\label{eq:eps_approx_app}
    \epsilon_\text{approx}=\frac{\left|\Omega\right|^2}{\Delta^2}\frac{\gamma}{\kappa}\frac{\Delta_a^{\text{opt}}}{\Delta}
\end{equation}
if the condition in Eq.~\eqref{eq:optimal_det} for the optimal detuning is satisfied, hence introducing a new source of error in the total infidelity.

Hence, what Fig.~\ref{fig:appB_coop} shows is that for high values of $\gamma/\kappa$ (Fig.~\ref{fig:appB_coop}(a)) the dominant source of error is due to the adiabatic elimination procedure, $\epsilon_{\text{approx}}$, as shown in Eq.~\eqref{eq:eps_approx_app}, so the infidelity is independent of the cooperativity. If this ratio is reduced, the error due to the losses $\epsilon_{\text{losses}}$ starts playing a role (Fig.~\ref{fig:appB_coop}(b)), but the scaling of the total error as $\sim 1/\sqrt{C}$, as predicted by Eq.~\eqref{eq:epsilon_losses_app}, is only possible if $\gamma/\kappa$ is small enough to neglect the error due to the projection onto the slow subspace, as shown in Fig.~\ref{fig:appB_coop}(c).

\begin{figure}[h]
    \centering
    \includegraphics[width=0.8\linewidth]{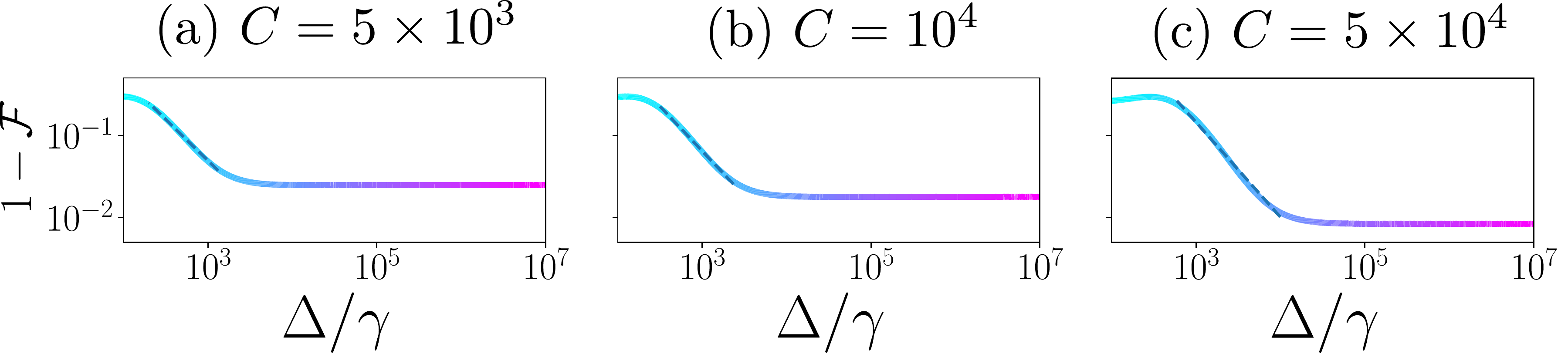}
    \caption{(a-c) Infidelity (Eq.~\eqref{eq:infidelity_def}) as function of $\Delta/\gamma$ for different values of the cooperativity $C$. The figures reproduce the results shown in the bottom panel of Fig.~\ref{fig:zz_agnostic}(c) of the main text, so the parameters taken are equal to the ones there with the except for the value of $C$. Furthermore, the linear fits superimposed over the loglog plots around the non-constant regions yield slopes of $-1.00$ (a), $-1.09$ (b) and $-1.15$ (c).}
    \label{fig:appB_gamma}
\end{figure}

Finally, in order to check the scaling of the total error as a function of the ratio $\Delta/\gamma$ for a fixed cooperativity discussed in the bottom panel of Fig.~\ref{fig:zz_agnostic}(c), we show in Fig.~\ref{fig:appB_gamma} a similar figure that in the main text (Fig.~\ref{fig:appB_gamma}(b) is indeed the same, shown here to compare) but including more values of the cooperativity. This plots show that the value in which the total error saturates is smaller as $C$ is increased, in agreement with the claim we made in the main text regarding this constant value as the effect of $\epsilon_{\text{losses}}$, which is smaller as the cooperativity increases.

\end{widetext}

\end{document}